\documentclass[twocolumn]{aastex631}

\usepackage{amsmath} 
\graphicspath{{./}{figures/}}

\usepackage{hyperref}

\newcommand\oms{\Omega_{\rm S}}
\newcommand\omd{\Omega_{\rm dyn}}
\newcommand\psat{2\pi/\oms}
\newcommand\req{R_{\rm eq}}
\newcommand\depth{d/\req}
\newcommand\rsurf{r_{\rm surf}}
\newcommand\omp{\Omega_{\rm p}}
\newcommand\ompi{\Omega_{{\rm p},i}}
\newcommand\rout{r_{\rm out}}
\newcommand\zout{Z_{\rm out}}
\newcommand\zin{Z_{\rm in}}
\newcommand\yout{Y_{\rm out}^\prime}
\newcommand\yin{Y_{\rm in}^\prime}
\newcommand\yp{Y^\prime}
\newcommand\degd{{\rm deg\ d}^{-1}}

\newcommand\fice{f_{\rm ice}}

\shorttitle{Saturn winds and seismology}
\shortauthors{Mankovich, Dewberry, Fuller}

\begin{document}

\title{Saturn's Seismic Rotation Revisited}

\author{Christopher R. Mankovich}
\affiliation{Division of Geological and Planetary Sciences, Mailcode 150-21, California Institute of Technology, Pasadena, CA 91125, USA; \url{cmankovich@ucsc.edu}}

\author{Janosz W. Dewberry}
\affiliation{TAPIR, Walter Burke Institute for Theoretical Physics, Mailcode 350-17, California Institute of Technology, Pasadena, CA 91125, USA}
\affiliation{Canadian Institute for Theoretical Astrophysics, 60 St. George Street, Toronto, ON M5S 3H8, Canada}

\author{Jim Fuller}
\affiliation{TAPIR, Walter Burke Institute for Theoretical Physics, Mailcode 350-17, California Institute of Technology, Pasadena, CA 91125, USA}

\begin{abstract}
Normal mode seismology is a promising means of measuring rotation in gas giant interiors, and ring seismology presents a singular opportunity to do so at Saturn. We calculate Saturn's normal modes of oscillation and zonal gravity field, using nonperturbative methods for normal modes in the rigidly rotating approximation, and perturbative methods for the shifts that Saturn's deep winds induce in the mode frequencies and zonal gravity harmonics. The latter are calculated by solving the thermo-gravitational wind equation in an oblate geometry. Comparing many such models to gravity data and the frequencies of ring patterns excited by Saturn normal modes, we use statistical methods to estimate that Saturn's cloud-level winds {extend inward along cylinders before decaying} at a depth $0.125\textrm{--}0.138$ {times Saturn's equatorial radius}, or $7,530\textrm{--}8,320$ km, consistent with analyses of Cassini gravity and magnetic field data. The seismology is especially useful for pinning down Saturn's poorly constrained deep rotation period, which we estimate at $\psat=634.7$~min (median) with a 5/95\% quantile range $633.8-635.5$~min. Outstanding residuals in mode frequencies at low angular degree suggest a more complicated deep interior than has been considered to date. Smaller but still significant residuals at high angular degree also show that our picture for the thermal, composition, and/or rotation profile in Saturn's envelope is not yet complete.
\end{abstract}

\section{Introduction} \label{sec.intro}
Precision gravity science from the Juno spacecraft at Jupiter and Cassini at Saturn has provided evidence for deep differential flows in the atmospheres of both planets \citep{2018Natur.555..223K,2018Natur.555..227G,2019Sci...364.2965I,2019GeoRL..46..616G,2019ApJ...879...78M} and has stimulated new attention toward the heavy element gradients that may be a common feature of gas giant interiors \citep{2017GeoRL..44.4649W,2019ApJ...872..100D,2020AREPS..48..465S,2021PSJ.....2..241N,2022A&A...662A..18M,BM22}. But much is still unclear about these planets' internal structures. In the case of Saturn, the near-perfect alignment between the rotation axis and the magnetic dipole axis \citep{2020Icar..34413541C} obscures the spin of Saturn's deep interior, a key input for structure models.
Hence, alternative observational means of disentangling Saturn's rotation and composition structure are highly sought after.

Parallel to the gravity science, 
{Juno and Cassini have also renewed interest in giant planet seismology. Anomalous spacecraft accelerations appear to support the idea of acoustic (p-)mode seismicity in Jupiter \citep{2022NatCo..13.4632D} and Saturn \citep{2020PSJ.....1...27M}, although this offers no immediate quantitative constraints {on planetary structure}. On the other hand,}
the unique data set of ring seismology at Saturn has proven useful for constraining a giant planet's interior in new ways. Spurred on by observations by \cite{1991Icar...93...25R} and \cite{2011Icar..216..292B} and detailed predictions laid out by \cite{1991Icar...94..420M} and \cite{1993Icar..106..508M}, a growing corpus of Cassini detections of Saturn-associated ring waves \citep{2013AJ....146...12H,2014MNRAS.444.1369H,2016Icar..279...62F,2019Icar..319..599F,2019AJ....157...18H,2021Icar..37014660F,2022PSJ.....3...61H} has provided precise measurements of the frequencies of 34 ring waves that can each be traced back to a distinct {nonradial} normal mode of oscillation inside Saturn. 

Most of these waves take the form of spiral density waves at outer Lindblad resonances (OLRs) or spiral bending waves at outer vertical resonances (OVRs)\footnote{In addition to these 34, \cite{2021Icar..37014660F} report two Saturn-associated inner Lindblad resonances (ILRs), but these stem from the same Saturn modes responsible for known OLRs, and hence do not add new frequency constraints.}. The corresponding Saturn modes can {typically}\footnote{{
Degree $\ell$ and radial order $n$ are not  good quantum numbers for modes strongly affected by rotation or mode-mode degeneracies. The Saturn modes at hand take the shape of slightly distorted spherical harmonics and hence can still be assigned a single $\ell$ describing the \emph{dominant} angular degree. Radial order $n$ can lose its meaning entirely in cases involving degeneracy, e.g., in rosette modes, {which are} not the focus of this work.
}}
be categorized by their azimuthal order $m$, dominant angular degree $\ell$, and number of radial nodes $n$.
The known Saturn-associated detections are prograde modes spanning azimuthal orders $m=2-14$, and comparison to models \citep[][``M19'']{2019ApJ...871....1M} makes it clear that these data correspond to Saturn modes with $\ell-m=0-8$ \citep{2021Icar..37014660F}. 
{(Because we deal here with exclusively prograde patterns, we use the convention that $m>0$ denotes prograde planet modes and ring patterns.)}
Most of the detected modes are the fundamental (f-)modes with $n=0$, effectively surface gravity modes that are confined increasingly close to the planet's surface with increasing $\ell$. M19 found the f-modes to be highly sensitive to Saturn's rotation, using the 14 f-mode frequencies that were known at the time to make the first seismic estimate of Saturn's bulk rotation period. 

Beyond the f-modes, the observations have also uncovered internal gravity (g-)modes with $n\geq1$ at the lowest azimuthal orders $m=2,3$, where they can induce finely separated mode multiplets via degenerate mixing with f-modes \citep{2014Icar..242..283F}, or simply appear as isolated g-modes \citep[][``MF21'']{2021NatAs...5.1103M}. This striking result of ring seismology directly implies that not all of Saturn's interior is convective, as the classical picture would have it. MF21 identified an instance of an isolated g-mode in the $m=2$ density wave W76.44, ascribing it to the $_{\,\ell}^m\cdot_n=\,_2^2g_1$ mode and using its frequency to constrain the extent of the stably stratified g-mode cavity associated with Saturn's diffuse core. Other isolated low-order g-modes may be observable through $m\lesssim5$. \citet[][``D21'']{2021PSJ.....2..198D} addressed the phenomenon of mode mixing and fine-scale multiplicity in detail, considering the additional influence of the differential rotation implied by the Cassini gravity results. They found that Saturn's deep winds affect the frequencies of f-modes, which in turn can drive mixing with other modes by bringing their frequencies toward resonance.

Critically, while \cite{2014Icar..242..283F}, M19, and MF21 used perturbation theory to account for Saturn's rapid rotation in the oscillation calculations, D21 employed a nonperturbative method for solving for modes in a two-dimensional oblate geometry \citep[see, e.g.,][]{2006A&A...455..621R,2012A&A...547A..75O,2017PhRvD..96h3005X}, the first time such a calculation has been applied to Saturn ring seismology. This technical leap led to the identification of a host of `rosette' modes, modes that are essentially coupled {sequences} of g-modes strongly influenced by rotation{, named for their} rosette-shaped eigenfunctions (see D21, Figures 3-5; see also \citealt{2012ASPC..462..389B} and \citealt{2013PASJ...65...68T}).

More germane to the question of Saturn's rotation state is the major accuracy advantage of nonperturbative methods compared to the approximate nature of perturbative methods: The latter (e.g., M19) have generally included effects to second order in the ratio $\oms/\omd$ where $\oms$ is spin frequency and $\omd\equiv(GM_{\rm S}/R_{\rm eq}^3)^{1/2}$ is Saturn's equatorial dynamical frequency. For Saturn, $\oms/\omd\sim0.4$ so that mode frequencies from second-order theories are uncertain at the level of $(\oms/\omd)^3\sim5\%$, a major limitation considering that ring observations yield frequencies to precisions better than $0.01\%$ \citep{2021Icar..37014660F}. A realistic model certainly must also abandon the rigid rotation approximation and address Saturn's deep winds, which at $|\delta\Omega/\oms|\sim1\%$ may modify mode frequencies at the level of $\delta\Omega/\omd\sim0.4\%$. In contrast to the {rapid} bulk rotation, the rather weak differential rotation lends itself well to a perturbative treatment in mode calculations, an approximation whose validity \cite{D22} quantified by comparing with fully nonperturbative calculations for differentially rotating models.

These theoretical tools and the clearer understanding of Saturn's deep wind structure warrant a reappraisal of the seismic rotation rate inferred by \cite{2019ApJ...871....1M}. Here we use the zonal gravity field and frequencies of f-modes observed in the rings to jointly constrain the flow profile in Saturn's envelope, making the common assumption that the flow at depth follows the {alternating eastward and westward zonal jet streams} observed at the cloud level, with a simple decay prescription as a function of depth. 
From Markov chain Monte Carlo (MCMC) analysis we find Saturn's wind depth and deep spin to both be consistent with previous estimates, and use precision ring seismology through a nonperturbative numerical method to place a {more precise} constraint on the deep spin than has been possible with approximate perturbative methods.

Section~\ref{sec.methods} describes our interior model and shape calculation, our method for computing modes in the presence of rapid rotation and their frequency shifts arising from differential rotation, the wind-induced contributions to zonal gravity, and finally our MCMC sampling procedure. We present our findings in Section~\ref{sec.results}, discuss in  Section~\ref{sec.discussion}, and summarize in Section~\ref{sec.conclusion}.

\section{Methods} \label{sec.methods}
In brief our method is as follows: we
\begin{enumerate}
    \item Build rigidly rotating models for Saturn's oblate background structure and even zonal gravity field $J_{2n}$.
    \item Calculate each model's adiabatic oscillation modes (and frequencies $\sigma_i$) including non-perturbative treatment of rotational forces, considering many azimuthal orders $m$.
    \item Prescribe a profile for the decay of Saturn's surface winds toward the planet's center.
    \item Use a model based on {geostrophic (thermal wind)} balance to estimate the density perturbations and resulting shifts $\delta J_{n}$ to the zonal gravity stemming from the differential rotation.
    \item Perform a perturbative calculation of the frequency shift each mode attains as a result of the differential rotation.
    \item Finally evaluate each model against the observed $J_{2n}^{\rm obs}=J_{2n}+\delta J_{2n}$ and ring wave pattern speeds $\ompi=\sigma_i/m$ through a multivariate normal likelihood function $L$. A Markov Chain Monte Carlo algorithm samples interior models from this likelihood function, providing posterior distributions of the model parameters of interest.
\end{enumerate}
Each of these steps is detailed in the following.

\subsection{Rigidly rotating background} \label{sec.methods.rigidmodel}
We begin with rigidly rotating models for Saturn's interior structure. 
The initial step of solving for the rotating hydrostatic equilibrium with prescribed compositional and thermal structure and tabulated equations of state (EOS) is an iterative process, making the fully two-dimensional methods of D21 an impractical starting point. Instead we start by solving for the oblate equilibrium structure using a {seventh}-order perturbative theory of figures ({ToF7; \citealt{1978ppi..book.....Z,2021PSJ.....2..241N}}). The material is modeled as a mixture of H {(mass fraction $X$)}, He {(mass fraction $Y$)}, silicates, and ices by combining the ``MH13-SCvH'' EOS for H/He \citep{2013ApJ...774..148M,2016A&A...596A.114M} with the ANEOS \citep{aneos} EOS for serpentine and water ice. 
Constituent species are combined assuming an idealized mixture of noninteracting species, such that intensive quantities like the specific volume $1/\rho$ are sums over that of each species, weighted by the species' mass fractions. Note however that nonideal contributions to the entropy of H/He mixtures are inherently included in the \emph{ab initio} based \cite{2013ApJ...774..148M} EOS.

We take Saturn's composition structure to include a stably stratified dilute core, the simplest interior structure permitted by the low degree ring seismology (MF21, D21). We take the further step of assuming that this gradient extends to Saturn's center, leaving its outer radial extent $\rout$ as a free parameter such that the heavy element mass fraction follows
\begin{equation}\label{eq.z_profile}
    Z(r) = \left\{
    \begin{array}{l}
    \zout + (\zin-\zout)\sin^2\left[\frac\pi2\left(1-\frac{r}{\rout}\right)\right], \\
    \hfill r < \rout; \\
    \zout, \hfill r\geq \rout.
    \end{array}
    \right.
\end{equation}
While MF21 could not rule out the existence of a small ($\lesssim\!5\ M_{\rm E}$) chemically homogeneous, adiabatically or sub-adiabatically stratified inner core, much of the constraining power in the present analysis is in the high-degree ($\ell\gtrsim5$) f-modes whose eigenfunctions are localized close to Saturn's surface ($r/R_{\rm eq}\gtrsim0.8$) and so the details of the inner core are not crucial.
Finally, in lieu of direct knowledge of Saturn's helium distribution, we presume as in MF21 that the helium abundance is stratified over the same dilute core region such that an equation analogous to Eq.~\ref{eq.z_profile} holds with $\yp\equiv Y/(X+Y)=Y/(1-Z)$ appearing in place of $Z$ and $\yin$ and $\yout$ defined analogous to $\zin$ and $\zout$. This parameterization allows for a range of possible helium profiles, from homogeneous to bottom-heavy distributions consistent with hydrogen-helium phase separation, the quantitative understanding of which is still greatly uncertain \citep{2020NatRP...2..562H,2021Natur.593..517B}.

The condition that the mean of $Y^\prime$ over Saturn's volume satisfies the protosolar helium abundance $Y^\prime_{\rm proto}=0.275\pm0.001$ (e.g., \citealt{2003ApJ...591.1220L,2009ARA&A..47..481A}) eliminates $\yin$. {Tests treating $Y^\prime_{\rm proto}$ as a free parameter confirm that this uncertainty, here estimated from the range of available literature values, bears negligibly on our findings.}
Fixing the equatorial radius at $P=1\ {\rm bar}$ to 60,268 km \citep{1985AJ.....90.1136L} and demanding that the model's total mass satisfies Saturn's total mass in turn eliminates $\zin$. 
{We assume that the \emph{thermal} stratification is adiabatic throughout Saturn's interior, including the diffuse core, and fix} the temperature at $P=1\ {\rm bar}$ to be $T_1=135\ {\rm K}$ \citep{1985AJ.....90.1136L}{. Although the diffuse core likely has a superadiabatic temperature gradient associated with double-diffusive convection, its effect on fluid stability is overwhelmed by that of the composition gradient (MF21) and is hence negligible in the present analysis. W}e are left with a handful of free parameters $\zout,\ \yout,\ \rout,$ and the angular spin frequency $\oms$ of the rigidly rotating deep interior.

\begin{figure*}[ht]
\begin{center}
\includegraphics[width=\textwidth]{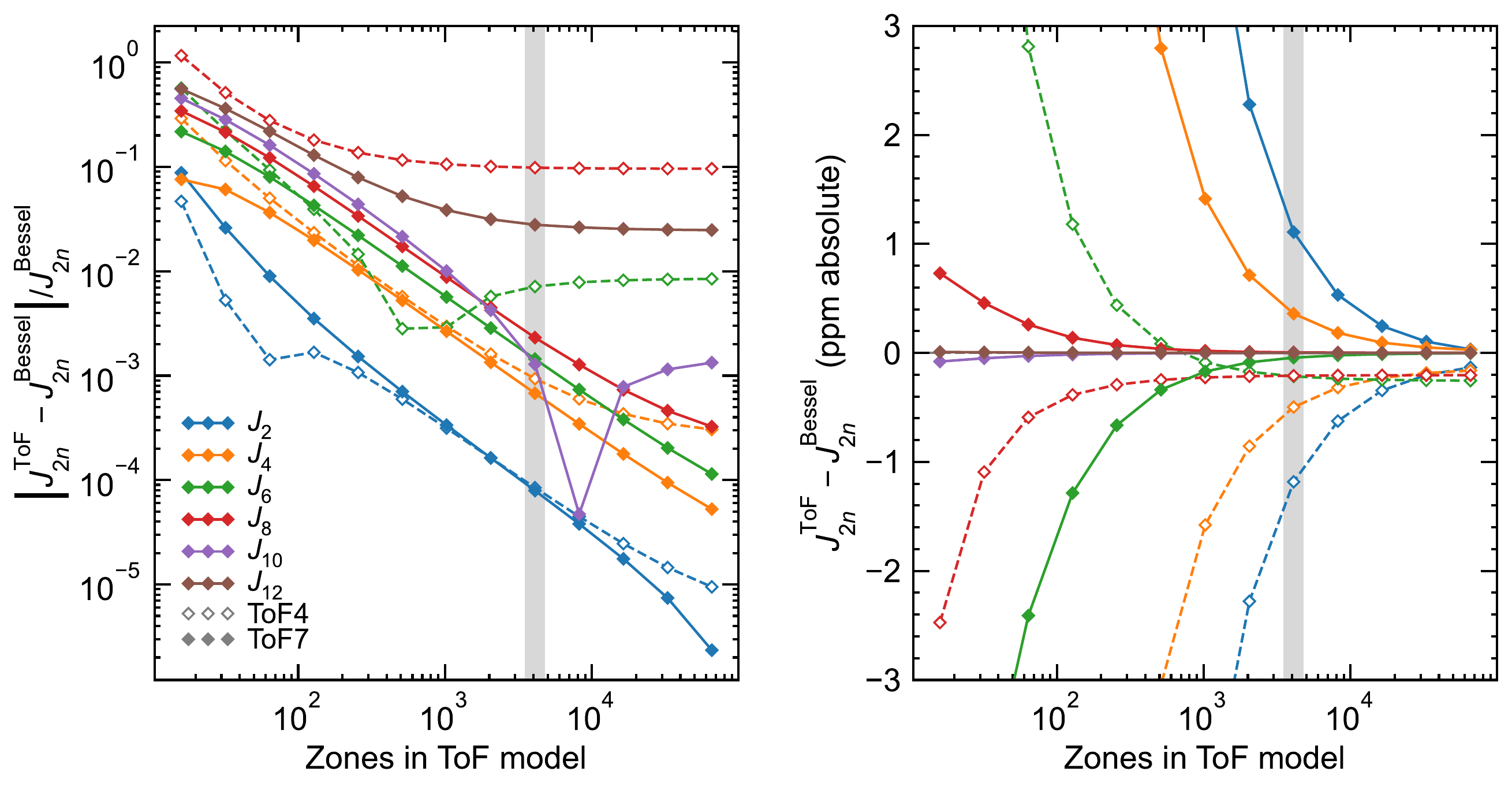}
\caption{\label{fig.tof_n1poly_error} Errors in zonal gravity harmonics for the $n=1$ polytrope at $q\equiv(\Omega^2 R_{\rm eq}^3/GM)_{\rm Jup}=0.089195487$, compared to the exact Bessel solution given by \cite{2016Icar..267..315W}. {Open symbols indicate ToF4 results and filled symbols indicate ToF7 results.} {ToF4 only yields even coefficients up to $J_8$.}
{Shaded regions} indicate our working resolution of 4,096 level surfaces and the corresponding $J_{2n}$ errors {in ToF7} dictate the variances adopted in the gravity likelihood function $L_{\rm grav}$ (see Eq.\ref{eq.likelihood}).
}
\end{center}
\end{figure*}

The fidelity of the calculated rigid-body $J_{2n}$ is limited by the accuracy of our {ToF7} implementation, which we benchmark using an $n=1$ polytrope for which a precise solution for the $J_{2n}$ is known \citep{2016Icar..267..315W,2021PSJ.....2..241N}. The resulting errors are shown as a function of spatial resolution in Figure~\ref{fig.tof_n1poly_error}. From the results of the benchmark we conclude that 4,096 level surfaces offers an acceptable accuracy, and adopt the corresponding fractional errors $(\Delta J_2/J_2, \Delta J_4/J_4, \Delta J_6/J_6)=(8\times10^{-5},7\times10^{-4},1\times10^{-3})$ as our effective uncertainties. These errors, significantly larger than the measurement uncertainties \citep{2019Sci...364.2965I,2022AJ....164..199J}, are similar to the residuals found at comparable resolution in the independent {ToF7} implementations of Nettelmann and Movshovitz \citep{2021PSJ.....2..241N}. 

{Although Figure~\ref{fig.tof_n1poly_error} shows typically one-sided errors for each $J_{2n}$, we find that in general, these errors can be positive or negative depending on the particular model in question. 
Hence, we conservatively treat the $\Delta J_{2n}/J_{2n}$ observed from this figure as symmetric (two-sided) uncertainties.}

\subsection{Solving for normal modes}
\label{sec.methods.modes}
Normal modes are first computed for the rigidly rotating background using the method described by D21 that includes a fully nonperturbative treatment for the effects of the Coriolis force and centrifugal distortion arising from Saturn's rapid rotation. We refer to that work for the full details of the calculation. These pseudospectral calculations employ nonspherical coordinates $(\zeta,\mu=\cos\theta)$ with $2N_\mu+1$ grid points in latitude and $N_\zeta$ grid points in the quasi-radial coordinate $\zeta${; see Appendix A in D21 for an explicit definition of these coordinates}. The density $\rho$, pressure $P$, Brunt-V\"ais\"al\"a (buoyancy) frequency $N$, and adiabatic sound speed $c_s$ are interpolated from the ToF model onto the much lower radial resolution oblate model used for the mode calculations.  {Aiming to balance frequency fidelity against computational expense, w}e find that a rather modest radial resolution $N_\zeta=40$ is sufficient for pattern speeds of the relevant modes to be converged to within 0.1 $\degd$ (a fractional frequency error $\delta\sigma/\sigma\lesssim0.01\%$), matching a conservative estimate of the precision on measured ring pattern speeds \citep{2021Icar..37014660F}. Computation time scales less steeply with $N_\mu$, which we fix at 61. Finally, spectral convergence is critical: rotation endows any equatorially symmetric mode\footnote{The same truncation parameter $N_{\tilde\ell}$ applies to antisymmetric modes, which get contributions from $\tilde\ell=m+1,m+3,m+5,\ldots$.} of a given $m$ value with contributions from angular degrees $\tilde\ell=m,m+2,m+4,\ldots$ and the number $N_{\tilde\ell}$ of terms retained in the spectral expansion must be chosen judiciously. We find that for high-degree f-modes, up to 12 values of $\tilde\ell$ contribute to an extent that affects the frequencies at $\delta\sigma/\sigma\lesssim0.01\%$. Hence when searching a given value of $m$ and modes up to a given $\ell-m$ are desired, retaining $N_{\tilde\ell}=12+\ell-m$ terms is sufficient. {The lower degree ($\ell\leq5$) modes that we fit in this work are more spectrally rich and require $N_{\tilde\ell}=20-24$.}
{In what follows, the modes under discussion always have an unambiguous \emph{dominant} angular degree that we label simply as $\ell$.}

\begin{figure*}
\begin{center}
\includegraphics[width=\textwidth]{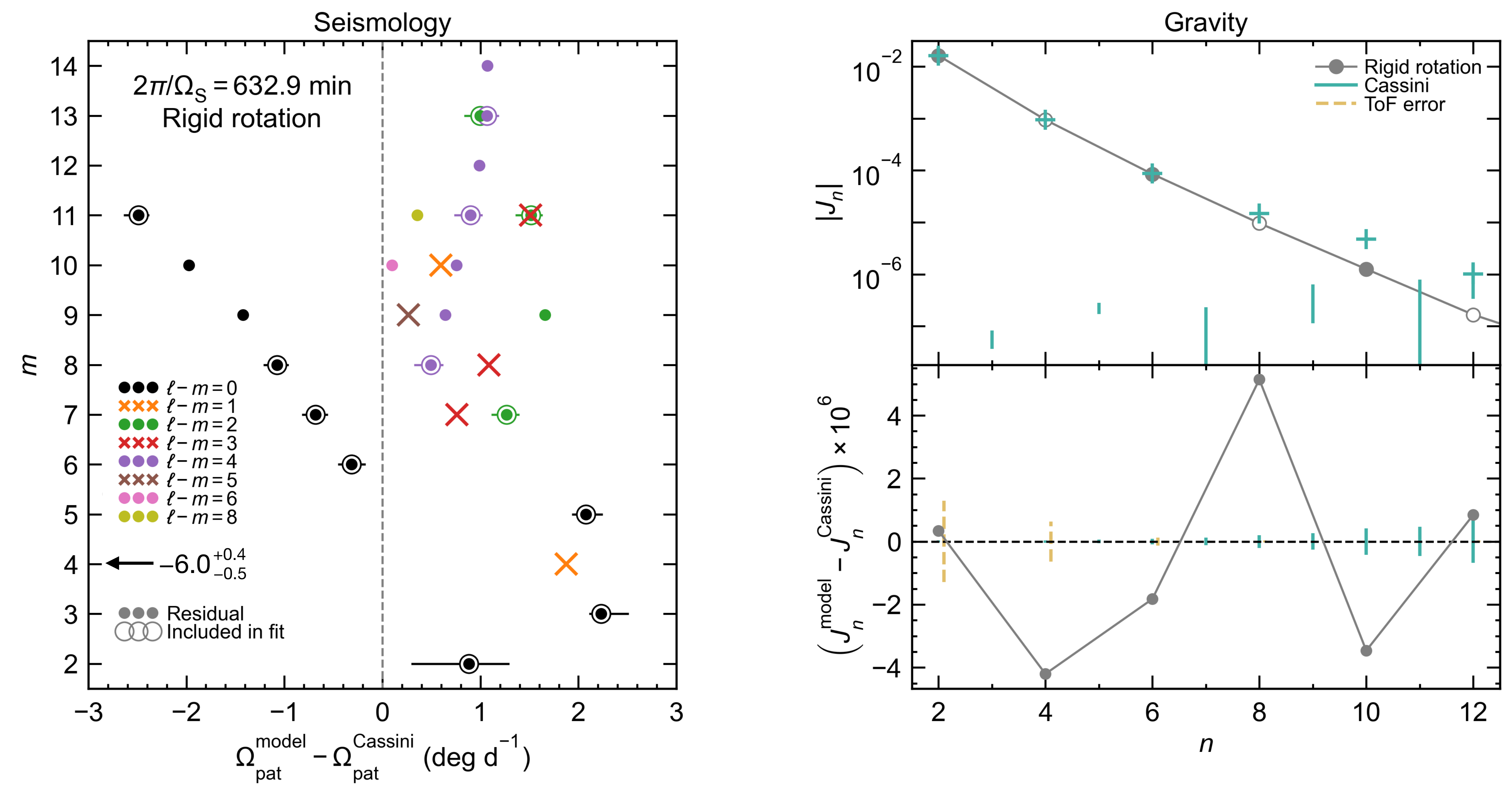}
\caption{
\label{fig.audit_rigid} 
\emph{Left:} Residuals (model minus data) of C ring density and bending wave pattern speeds for the best fitting rigidly rotating Saturn model. Different Saturn modes are distinguished by their azimuthal wavenumber $m$ (vertical axis) and latitudinal wavenumber $\ell-m$ (see colors in legend). Solid circles are fully nonperturbative calculations for mode frequency; open circles highlight modes calculated on the fly and fit to data during MCMC sampling (see Section~\ref{sec.methods.mcmc}). Error bars on these modes indicate 5-95\% quantiles of their residuals across the sample of rigid models. The systematic offsets seen for this optimized rigid model, most obvious for the sectoral ($\ell-m=0$) modes, betray the influence of Saturn's deep winds.
\emph{Right:} Gravitational moments $J_{2n}$ of the rigid model compared to data. Top panel plots the negative $J_{2n}$ as open circles, positive coefficients as filled circles. Bottom panel plots their absolute residuals with respect to the observed $J_{2n}$ \citep{2019Sci...364.2965I}.
}
\end{center}
\end{figure*}

Because our present emphasis is on Saturn's rotation profile rather than deep interior structure, and we require a lightweight enough calculation to efficiently sample a large number of models, we sidestep {some of} the lowest $m$ modes where multiplicities of observed ring waves arise from mixtures of fundamental and gravito-inertial modes (\citealt{2014Icar..242..283F},  MF21, D21). 
A complete description of this phenomenon arising in the $m=2$ and $m=3$ wave detections may require more than one g-mode cavity in the interior.
Instead we limit real time mode calculations to 
{f-modes with $m\geq5$, where all Saturn-associated ring waves appear to be excited by well isolated f-modes residing mostly in the envelope (M19, \citealt{2021Icar..37014660F}). 
To these we add a single deep-seated mode at each of $m=2,3,4$. At $m=3$ and $m=4$ these are $\ell=m$ f-modes that interact somewhat with nearby g-modes. 
The more nuanced case of $m=3$ involves three closely spaced detections that likely arise from interactions between $_3^3f$ and nearby g-modes or rosette modes (\citealt{2014Icar..242..283F}, D21); in this work we simply compute $_3^3f$ and compare to the strongest $m=3$ wave W82.06.
At $m=2$ we focus on the} deep-seated mode $_2^2g_1$ that MF21 ascribed to the fastest $m=2$ density wave W76.44. 
As in MF21 this mode provides a crucial anchor for $N(r)$, mitigating otherwise major degeneracies between $\zout$, $\rout$, and $\yout$ seen when fitting only the zonal gravity field.
$_2^2g_1$ also has the major practical advantage of not mixing strongly with higher degree ($\ell\gg2$) g-modes, unlike other $m=2$ and $m=3$ waves where strong interactions between the $\ell=m$ f-modes and $\ell\geq m$ modes {may necessitate $N_\ell$ and $N_\zeta$ of order $100$, the expense of which puts these outside our present scope.}

\begin{table}
\centering
\begin{tabular}{lrcr}
\hline
Wave & Mode ($_{\,\ell}^m\cdot_n$) & Fit in MCMC & Reference\\
\hline
W76.44	& $_{2}^{2}g_1$   	& y 	& F19	\\ 
W82.06	& $_{3}^{3}f$     	& y 	& H19	\\ 
W74.93	& $_{5}^{4}f$     	& -	& F19	\\ 
W80.98	& $_{4}^{4}f$     	& y 	& H19	\\ 
W81.02a	& $_{5}^{5}f$     	& y 	& H19	\\ 
W81.43	& $_{6}^{6}f$     	& y 	& H19	\\ 
W81.96	& $_{7}^{7}f$     	& y 	& H19	\\ 
W76.46	& $_{9}^{7}f$     	& y 	& F19	\\ 
W74.67	& $_{10}^{\ 7}f$  	& -	& F19	\\ 
W82.53	& $_{8}^{8}f$     	& y 	& H19	\\ 
W76.24	& $_{11}^{\ 8}f$  	& -	& F19	\\ 
W74.51	& $_{12}^{\ 8}f$  	& y 	& F21	\\ 
W74.94	& $_{14}^{\ 9}f$  	& -	& F19	\\ 
W83.09	& $_{9}^{9}f$     	& -	& H19	\\ 
W79.04	& $_{11}^{\ 9}f$  	& -	& F21	\\ 
W76.02	& $_{13}^{\ 9}f$  	& -	& F19	\\ 
W83.63	& $_{10}^{10}f$   	& -	& H19	\\ 
W77.34	& $_{14}^{10}f$   	& -	& F21	\\ 
W81.87	& $_{11}^{10}f$   	& -	& F21	\\ 
W75.14	& $_{16}^{10}f$   	& -	& F21	\\ 
W78.51	& $_{15}^{11}f$   	& y 	& F21	\\ 
W81.02b	& $_{13}^{11}f$   	& y 	& H19	\\ 
W84.15	& $_{11}^{11}f$   	& y 	& F21	\\ 
W79.84	& $_{14}^{11}f$   	& -	& F21	\\ 
W74.76	& $_{19}^{11}f$   	& -	& F21	\\ 
W79.55	& $_{16}^{12}f$   	& -	& F21	\\ 
W82.61	& $_{15}^{13}f$   	& y 	& F21	\\ 
W80.49	& $_{17}^{13}f$   	& y 	& F21	\\ 
W81.33	& $_{18}^{14}f$   	& -	& F21	\\ 
\hline
\end{tabular}
\caption{
\label{tab.associations}
{
Associations between Saturn modes and resonant features observed in Saturn's C ring. F19: \cite{2019Icar..319..599F}; H19: \cite{2019AJ....157...18H}; F21: \cite{2021Icar..37014660F}.
}
}
\end{table}

For the model under consideration, many of the mode pattern speeds are strongly covariant, meaning that little information is gained from the arduous process of calculating and fitting the full set of observed modes for every model. Instead useful subsets are chosen to represent a diversity of angular degrees $\ell$ and latitudinal wavenumbers $\ell-m$, and models of interest are followed up with complete calculations. 

Figure~\ref{fig.audit_rigid} summarizes the best fit we find for a rigidly rotating Saturn fit to $J_2$, $J_4$, $J_6$, and the pattern speeds of 
{14 C ring waves. The full list of wave-mode associations used in this work, as well as the subset fit on the fly, is given in Table~\ref{tab.associations}.
}
The full model sampling process is detailed in Section~\ref{sec.methods.mcmc} below; here it suffices to say that this rigid model is for an optimized combination of $\zout$, $\rout$, $\yout$, and $\psat$. Following up on the full set of modes with corresponding density wave detections, we find pattern speed residuals ranging from $-2$ to $+3\ \degd$, amounting to an overall spread of approximately 50 times the fiducial observational uncertainty $0.1\,\degd$. Clearly rigid rotation is not an adequate description, even with a considerable amount of freedom in the composition structure and planetary spin.

The optimized rigid model's 632.9 min spin period agrees with the initial ring seismology estimate of $633.6^{+1.9}_{-1.3}\ {\rm min}$ reported by M19. 
But in contrast to that work, the improved accuracy of the nonperturbative method reveals significant systematic frequency residuals, most noticeable for the sectoral ($\ell-m=0$) modes, that betray the influence of Saturn's deep winds. The sectoral mode frequencies are generally underestimated relative to the others, suggesting a prograde differential rotation concentrated near the equator where the sectoral modes are predominantly confined. Furthermore this systematic underestimate gets worse with increasing angular degree $\ell$, notable because f-modes with higher $\ell$ are confined closer to Saturn's surface.
Together these outcomes point toward a prograde differential rotation confined mostly to the equator and strongest near Saturn's surface, consistent with the dominant feature of Saturn's zonal winds. 
From the perspective of the f-mode frequencies there is an inherent tradeoff between the depth or amplitude of this prograde feature and the unknown spin frequency of the deep interior, and hence in a complete model addressing Saturn's deep winds, we expect slower deep spin than in the rigid model shown in Figure~\ref{fig.audit_rigid} or in M19.

{The frequency residuals in Figure~\ref{fig.audit_rigid} differ from those previously obtained in M19 (their Figure 5) and \cite{2021Icar..37014660F} (their Figure 10) for several reasons. First, the new model is stably stratified at depth, fundamentally altering the spectrum by introducing g-modes and strongly affecting the frequencies of all $m\lesssim5$ modes. Second, the new model's improved treatment for rotation in the mode calculations affects all modeled frequencies, most strongly those with larger values of $m$, leading to clear differences compared to the perturbative calculations shown in M19 and \cite{2021Icar..37014660F}. The present model's improved $J_{2n}$ accuracy and inclusion of $J_6$ in the fit also contribute to the slightly faster bulk rotation favored here, driving the more obvious systematic trend in frequency residuals with $m$ than was seen in M19 or \cite{2021Icar..37014660F}. In any case, we emphasize that this optimized rigid model is a poor fit to seismology and gravity both; understanding Saturn's differential rotation is essential.
}

\begin{figure*}[ht]
\begin{center}
\includegraphics[width=\textwidth]{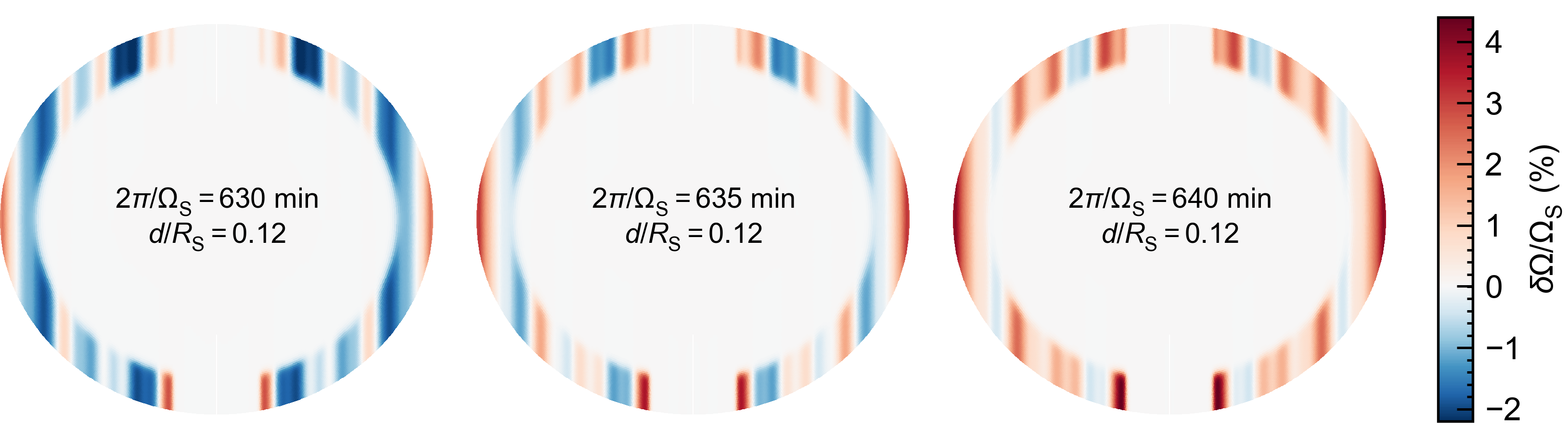}
\caption{\label{fig.rot_profiles_baroclinic} Example profiles of baroclinic differential rotation satisfying Saturn's observed cloud-level winds \citep{2011Icar..215...62G}, shown for fast, medium, and slow deep spin $\oms$ (left, center, and right columns) and with wind decay depth set to $\depth=0.12$. Depending on the deep spin, the winds can range from predominantly retrograde to predominantly prograde.}
\end{center}
\end{figure*}

\subsection{Baroclinic wind model}
\label{sec.methods.wind}
The winds observed at Saturn's surface are taken to extend inward on surfaces parallel to the rotation axis, before multiplication by a radial decay function centered at $r=\rsurf-d$ smoothly transitions to a rigidly rotating interior. In particular the spin frequency is given by $\boldsymbol\Omega(r,\mu)=\hat{\boldsymbol z}\left[\oms+\delta\Omega(r,\mu)\right]$ where
\begin{equation}\label{eq.decay}
\begin{split}
\delta\Omega(r,\mu) &= \delta\Omega_{\rm surf}(\tilde\mu) \\
&\times \frac12\left(1+\tanh\left[\frac{r-\left[r_{\rm surf}(\mu)-d\right]}{w}\right]\right).
\end{split}
\end{equation}
Here $\mu=\cos\theta$, $\rsurf=\rsurf(\mu)$ is the spherical radius coordinate of the oblate $P=1$ bar surface, and $\tilde\mu$ corresponds to the latitude on this surface sharing the same cylindrical radius as an interior point $(r,\mu)$. The depth parameter $d$ controls the centroid of the decay function, and $w$ is the decay width. 
Tests reveal that wind-induced perturbations to mode frequencies and gravity harmonics are only weakly sensitive to the decay width $w$; in what follows we fix $w=0.02$. 

{The assumed surface winds} $\delta\Omega_{\rm surf}(\mu)$ follow from a Legendre polynomial expansion of the {average} wind speeds observed {in the ISS narrowband continuum filters CB2/CB3} \citep{2011Icar..215...62G}, normalized to a given assumed deep rotation rate $\oms$. 
{We also consider a slightly modified surface wind profile (``Galanti reconstructed''; E. Galanti, private communication) whose main departure from the García-Melendo winds is its enhanced retrograde rotation in the region $0.4<|\mu|<0.6$.
This is anticipating that a rotation law of the form~\ref{eq.decay} and adhering strictly to the García-Melendo cloud-level velocity profile may have difficulty reproducing Saturn's observed odd and higher degree even ($J_{6+}$) gravity harmonics \citep{2019GeoRL..46..616G,2021MNRAS.501.2352G}.
}

{In both cases we find that a 60th order polynomial and our 123 zone latitudinal grid resolution (Section~\ref{sec.methods.modes}) are sufficient to resolve the $J_n$ and $\ompi$ to within our fit tolerances.}
{We retain both equatorially symmetric and antisymmetric terms in the expansion.
Although we will see that the antisymmetric components have \text{a negligible} effect on the frequencies of isolated f-modes, we do wish to consider the asymmetric gravity moments ($J_3,\ J_5,\ \ldots$) induced by the winds.
}
{In all cases wind speeds are forced to vanish} near the poles using an exponential decay at $|\mu|>0.975$, and antisymmetric terms are similarly attenutated close to the midplane to avoid discontinuities; for further details see D21.
Sample profiles of the differential rotation model are depicted in Figure~\ref{fig.rot_profiles_baroclinic}.

\subsection{Wind-induced perturbations to zonal gravity}
\label{sec.methods.djn}

{Saturn's winds modify the planet's equilibrium structure and gravity. We calculate shifts in the gravitational coefficients $\delta 
J_n$ by solving a ``thermal gravitational wind equation'' (TGWE; \citealt{2015ApJ...806..270Z,2017JGRE..122..686C}) that enforces 
geostrophic 
balance. \autoref{app:tgwe} describes the numerical approach we take to this problem. In short, we solve the coupled system of Euler and Poisson equations governing small-amplitude, time-independent density and gravity perturbations $\rho'({\bf r})$, $\Phi'({\bf r})$ that result from imposing a wind with a given decay profile. 

These calculations build on the common practice of solving the ``thermal wind equation'' (TWE) in a spherical geometry \citep{2019GeoRL..46..616G,2021MNRAS.501.2352G} with the inclusion of self-gravity and the already oblate background of the rigidly rotating model. Like \cite{2017JGRE..122..686C}, we do not find that self-gravity hugely affects the $\delta J_n$ for $n\geq 6$, but including it does allow for the simple imposition of the constraint that the total mass remain unchanged (i.e., that $\delta J_0=0$). On the other hand, surface oblateness induced by Saturn's rapid bulk rotation can have a larger effect, depending on the differential rotation profile (see \autoref{fig:tgwe}). 

Note that by construction both the TWE and TGWE neglect irrotational contributions to the Euler equation \citep{2017JGRE..122..686C}. While our comparisons with fully nonlinear calculations for differentially rotating polytropes (see \autoref{fig:tgwe}) suggest that the TGWE provides accurate $\delta J_n$ for $n\geq6$, fully nonlinear calculations of differentially rotating equilibria are desirable.} 

\begin{figure}
\includegraphics[width=\columnwidth]{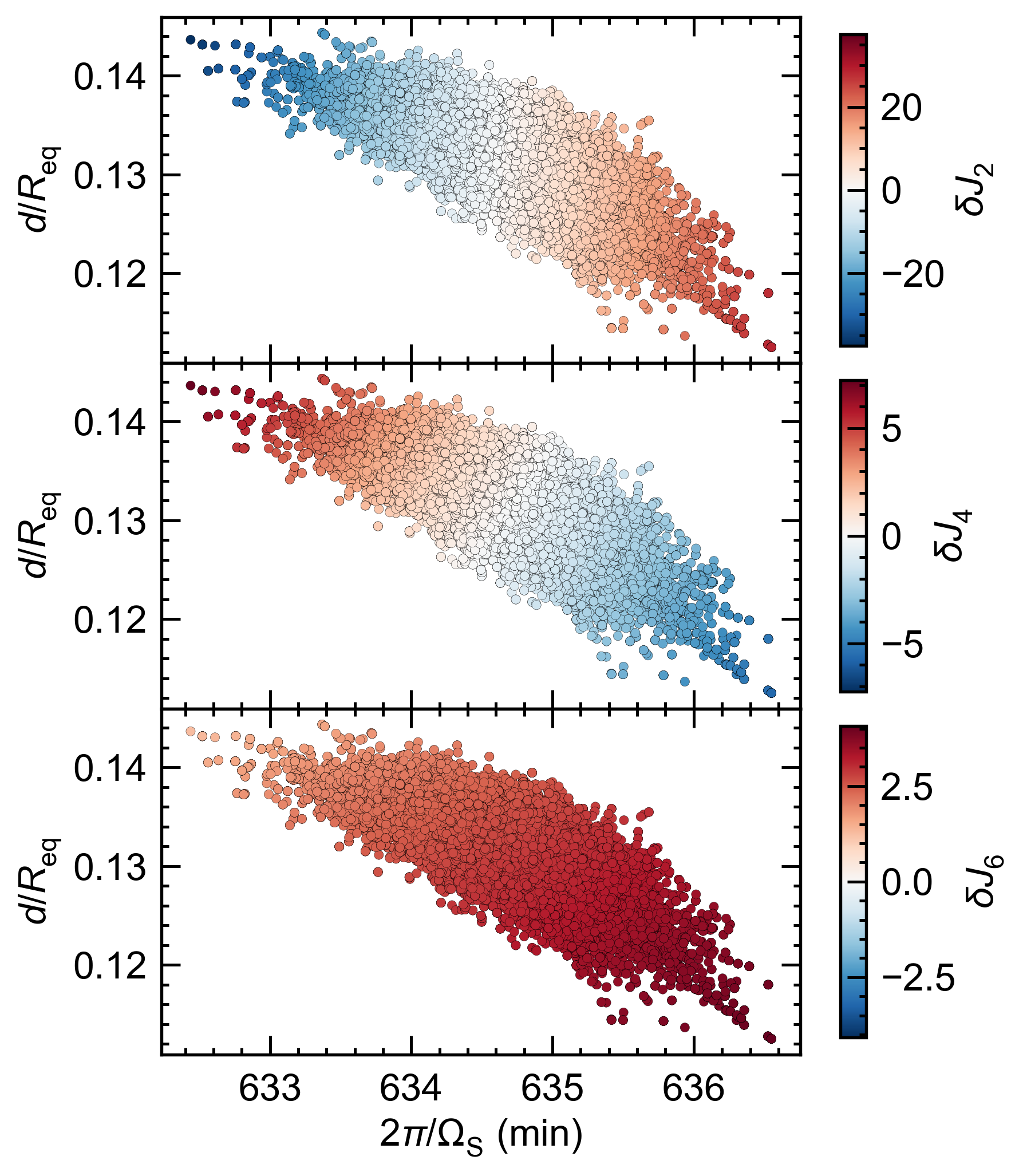}
\caption{
\label{fig.tgwe_oblate_djn_winds} 
Wind-induced zonal gravity perturbations $\delta J_{2n}$ calculated assuming thermal-gravitational wind balance for baroclinic rotation laws of the form {given in} Equation~\ref{eq.decay}, as a function of assumed deep spin and wind decay depth. These points are for our baseline sample that adopts a modified cloud-level wind (see Section~\ref{sec.methods.djn}). Color bars are labeled in parts per million.
}
\end{figure}

{Figure~\ref{fig.tgwe_oblate_djn_winds} summarizes the $\delta J_{2n}$ predicted by the oblate TGWE over the baseline sample of differentially rotating models that we obtain below. These models assume the Galanti reconstructed cloud-level winds. Winds enhance $J_6$ by typically $\delta J_6\approx3$ ppm.}

\subsection{Wind-induced perturbations to mode frequencies}
\label{sec.methods.dom}
\subsubsection{Method}

\begin{figure}
\includegraphics[width=0.9\columnwidth]{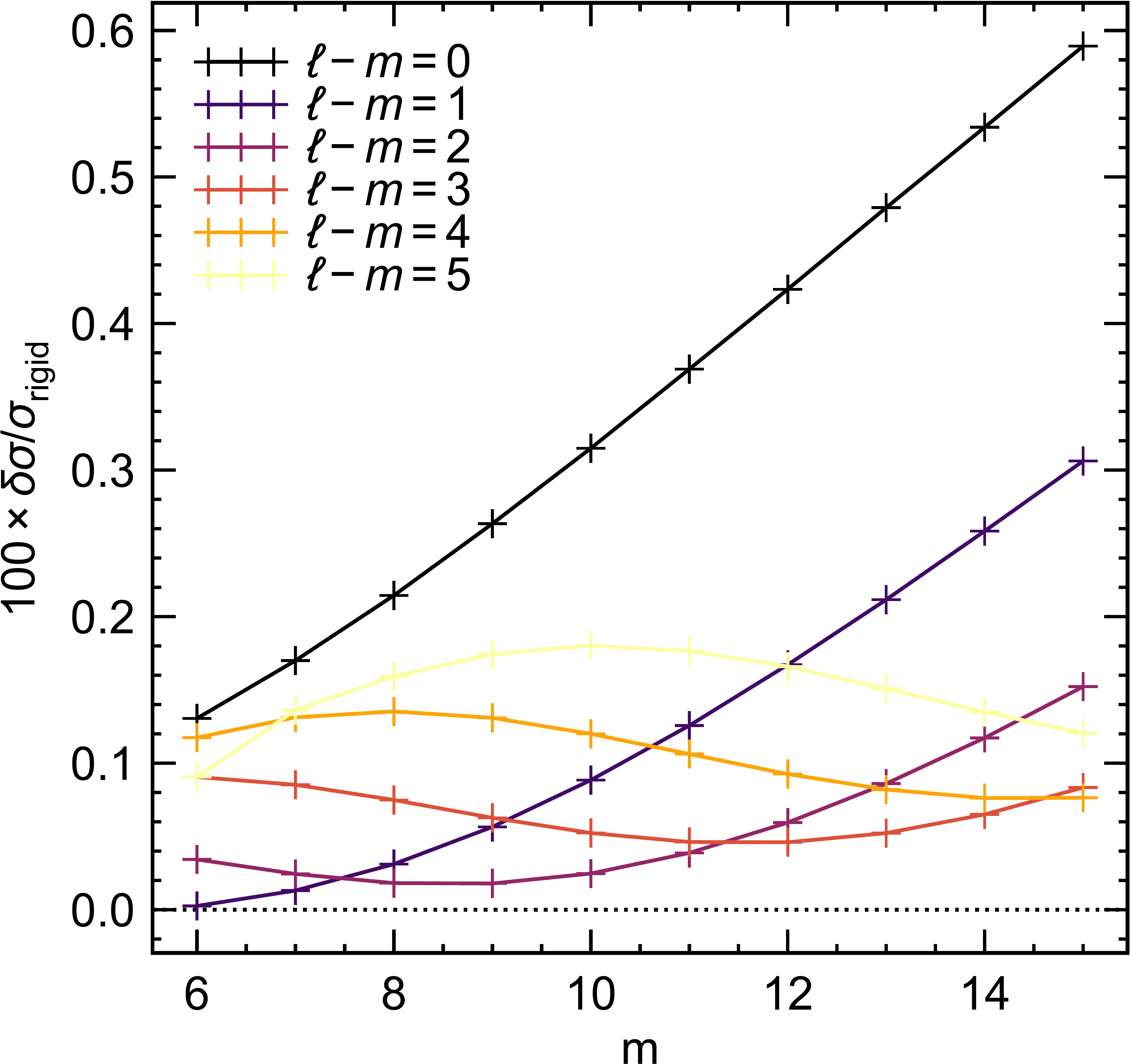}
\caption{
\label{fig.pert_freq_shifts} 
{Wind-induced perturbations to f-mode frequencies, calculated from Equation~\ref{eq.pertshift}, expressed here in terms of inertial frame frequency $\sigma_i=\omega_i+\delta\omega_i$. After Figure 4 in \cite{D22}.
}
}
\end{figure}

Saturn's $\lesssim300$ m s$^{-1}$ wind speeds at the cloud level translate to a modest differential rotation amplitude $|\delta\Omega/\oms|\lesssim3\%$. The order of magnitude influence on frequencies of oscillation modes is even smaller at $|\delta\Omega/\omd|\lesssim1\%$, motivating the use of a perturbation theory to calculate the leading order frequency shifts from differential rotation. We accordingly calculate these shifts accounting for the self-coupling induced in each mode by the winds, a combination of Coriolis and centrifugal effects, as {(D21 Equation 24)
\begin{equation}\label{eq.pertshift}
\delta\omega_i\simeq\frac{-(2\omega_i\mathcal W_{ii}^D+\mathcal U_{ii}^D)}{2(\omega_i\mathcal T_{ii}+\mathcal W_{ii}^R+\mathcal W_{ii}^D)}
\end{equation}
where $\omega_i$ is the frequency of the $i$th mode {under rigid rotation}, $\mathcal W_{ii}^D$ and $\mathcal W_{ii}^R$ are inner products involving the Coriolis operators associated with differential and rigid rotation respectively, $\mathcal T_{ii}$ corresponds to kinetic energy and is equal to the identity matrix in our chosen normalization, and $\mathcal U_{ii}^D$ corresponds to a potential energy operator involving cross terms between the rigid and differential parts of the rotation. D21 gives full integral expressions for these operators. Figure~\ref{fig.pert_freq_shifts} gives the frequency shifts calculated from Equation~\ref{eq.pertshift} for a subset of the $\ell\geq6$ modes in a representative model.}

This approach is accurate insomuch as (i) the frequency shifts are not large enough to cause accidental degeneracies between different modes with the same $m$, which D21 showed would lead to wind-induced mode mixing, and (ii) higher-order consequences of the differential rotation, such as feedback from wind-induced centrifugal flattening on the mode frequency, are negligible. Condition (i) is an excellent approximation for the high-degree f-modes, which as we have mentioned are extremely well isolated in frequency within a single $m$ family\footnote{The co-rotating f-mode frequencies scale roughly as $\omega_{\rm f}/\omd\sim\ell^{1/2}$ so that within a given $m$, each f-mode is separated from the next by at least $\Delta\omega_{\rm f}/\omd\sim(\ell+1)^{1/2}-\ell^{1/2}\approx20\%$ at $\ell=6$ and $10\%$ at $\ell=20$, safely an order of magnitude larger than the wind-induced shifts.}. 
Condition (ii) is more dubious, and with this question in mind \cite{D22} apply a complete nonperturbative treatment for Saturn-like differential rotation in an $n=1$ polytrope to conclude that the simpler perturbative approach adopted here and in D21 overestimates the wind-induced frequency shifts by up to 0.1-0.2\%. This small but significant offset is most noticeable for the highest degree sectoral ($\ell-m=0$) modes and can be attributed mostly to the perturbative method's neglect of the wind-induced modification to the equilibrium shape. 

Because fully nonperturbative calculations along the lines of \cite{D22} are not readily possible for the more realistic (non-polytropic) equation of state based models considered here,
{we caution that our perturbative calculation (\ref{eq.pertshift}) may slightly overestimate the true mode frequencies. 
We expect however that the frequency errors introduced by the perturbation theory will be smaller than in \citealt{D22}, mainly because of differences in the decay profile. Below we show that a suitable fit to the gravity field (particularly $J_6$) requires a modified cloud-level wind similar to that of \cite{2019GeoRL..46..616G} (see Section~\ref{sec.methods.djn}). This leads to a more strongly sub-corotating region at latitudes $0.4<|\mu|<0.6$ (Figure~\ref{fig.wind_eig_overlap_1d}), mitigating somewhat the effect of the predominant prograde equatorial jet. Hence we expect smaller frequency perturbations from the winds in general, especially for the tesseral ($\ell-m>0$) modes that sample higher latitudes. We also expect weaker feedback on frequencies from the wind-induced perturbation to the planetary shape, the predominant effect responsible for the perturbative/nonperturbative frequency offsets observed by \cite{D22}. Futhermore half the frequencies we fit on the fly correspond to $\ell-m>0$ modes, which \cite{D22} found are dramatically less affected by the perturbative approximation, their errors insignificant at $\delta\sigma/\sigma\lesssim0.05\%$.
}
{We finally note that our formulation takes Saturn's equatorial radius as the
reference length scale for non-dimensionalization, and hence excluding the shape change amounts to ignoring wind-induced displacement only at higher
latitudes.}

\begin{figure*}
\begin{center}
\includegraphics[width=\textwidth]{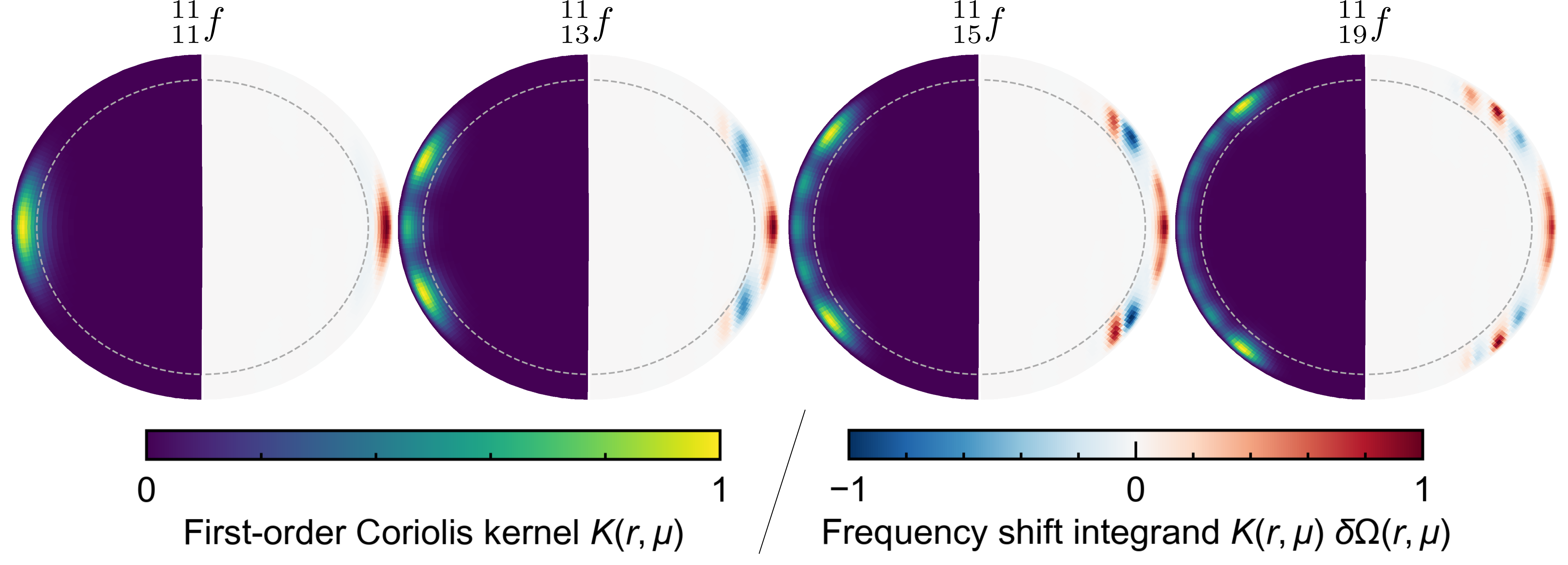}
\caption{
\label{fig.wind_eig_overlap} 
{Meridional slices} of the first-order Coriolis kernels (left hemispheres; Equation~\ref{eq.kernel}) and the resulting integrands of the first-order frequency shift (right hemispheres; Equation~\ref{eq.kernelshift}) for the four f-modes with $m=11$ and $\ell-m=0,\ 2,\ 4,\ 6$ 
in our favored Saturn model.
The dashed grey boundary marks the centroid of the wind decay function.
All four modes are shifted to higher frequency by the prograde equatorial jet, although modes with larger $\ell-m$ acquire smaller shifts because they sample higher latitudes where they suffer self-cancellation. Color bars are normalized arbitrarily; values of the frequency shift are given in the text. Note that the perturbative treatment of winds in our main calculations captures higher-order rotation effects than the heuristic first-order perturbations shown here.
}
\end{center}
\end{figure*}

\begin{figure*}
\begin{center}
\includegraphics[width=\textwidth]{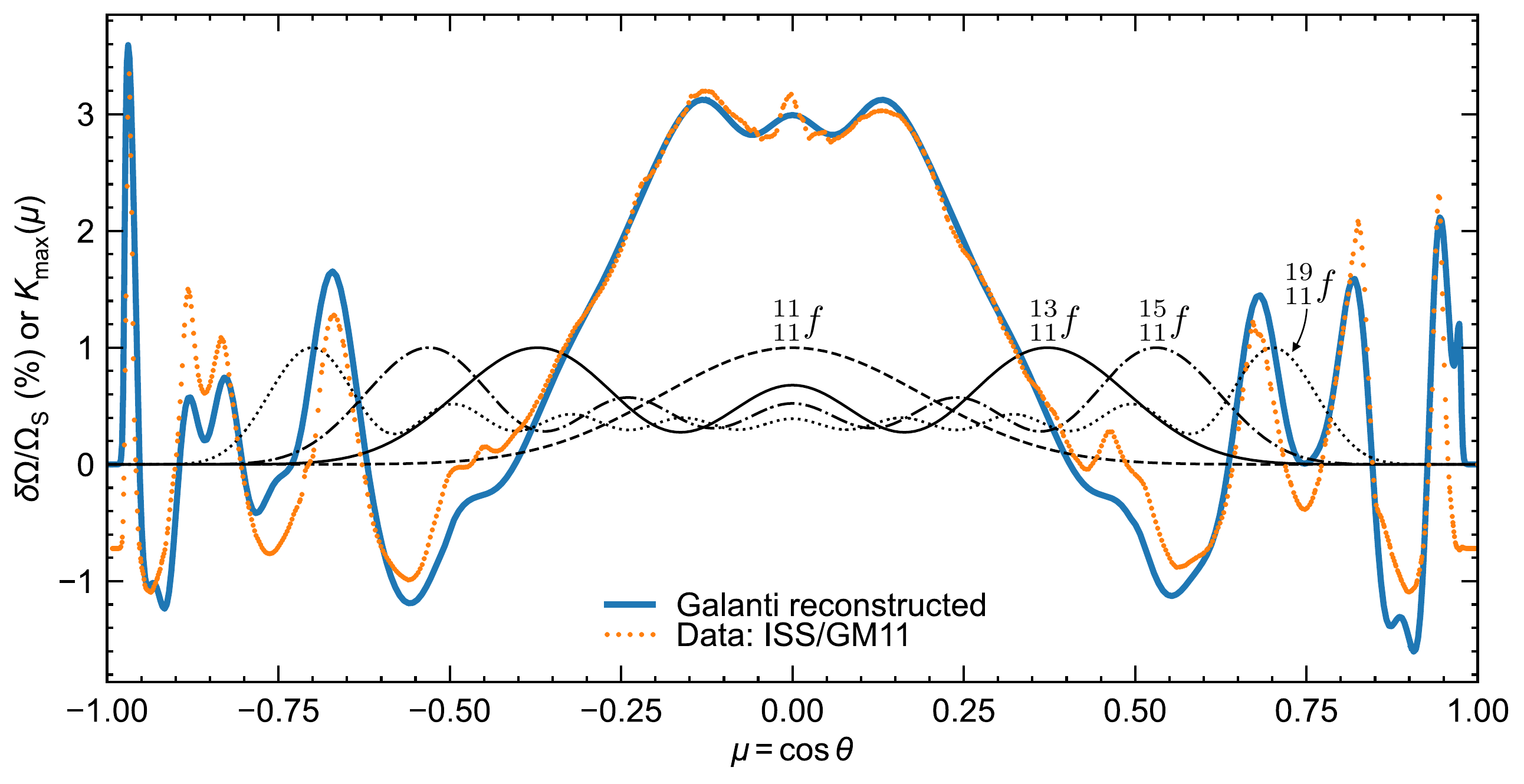}
\caption{
\label{fig.wind_eig_overlap_1d} 
Examples of a few modes' latitudinal sensitivity to rotation: as in Figure~\ref{fig.wind_eig_overlap}, but comparing the {Legendre expansion of the} cloud-level wind angular frequency with the radial peak amplitude reached by $K(r,\mu)$ (Equations~\ref{eq.kernelshift}-\ref{eq.kernel}) as a function of {planetocentric} latitude for each of the three modes (black curves with labels, normalized to unity). Small orange points are the mean measured wind profile from Casini imaging \citep{2011Icar..215...62G} and the solid blue curve is a gravity-optimized profile from \cite{2021MNRAS.501.2352G}; both are scaled to 635 min baseline rotation here. Modes with $\ell-m=0$ essentially only sample the prograde equatorial jet; $\ell-m=2$ and $4$ are more sensitive to the strongly retrograde regions from $30^\circ$ to $40^\circ$ latitude.
}
\end{center}
\end{figure*}

\subsubsection{Which modes probe which regions?}
{
A useful demonstration of the ring seismology's spatial sensitivity to Saturn's differential rotation can be gleaned from the perturbative shifts (Equation~\ref{eq.pertshift}).
Inspection of the inner products appearing in that equation for high-degree ($\ell\gtrsim5$) f-modes in a Saturn model shows that to a fair approximation, we can neglect $\mathcal U_{ii}^D$ relative to $2\omega_i\mathcal W_{ii}^D$ in the numerator, and neglect $\mathcal W_{ii}^R+\mathcal W_{ii}^D$ relative to $\omega_i\mathcal T_{ii}$ in the denominator, leaving
\begin{align}
\delta\omega_i&\simeq-\mathcal W_{ii}^D/\mathcal T_{ii} \\
&=\frac{\int\rho\,\delta\Omega\,\boldsymbol\xi_i^*\cdot\left(m\boldsymbol\xi_i-{\rm i}\hat{\boldsymbol z}\times\boldsymbol\xi_i\right)\,dV}{\int\rho\,\boldsymbol\xi_i^*\cdot\boldsymbol\xi_i\,dV}
\end{align}
with $\boldsymbol\xi_i$ denoting the Lagrangian displacement eigenfunction for the $i$th mode. 
Then the frequency shift can be succinctly written as (dropping the subscripts $i$)
\begin{equation}\label{eq.kernelshift}
\delta\omega\simeq\int\delta\Omega(r,\mu)K(r,\mu)\,dV
\end{equation}
by defining the first-order Coriolis kernel
\begin{equation}\label{eq.kernel}
    K=\frac{{\rho\left[m|\boldsymbol\xi|^2 - {\rm i}\boldsymbol\xi^*\cdot(\boldsymbol{\hat z}\times\boldsymbol\xi)\right]}}{\int\rho|\boldsymbol\xi|^2\,dV},
\end{equation}
a field independent of the assumed rotation model (and equivalent to, e.g., Equation 3.332 of \citealt{2010aste.book.....A}).
{Here we use this simplified rotation kernel for heuristic purposes; our main results will use the full shifts given by Equation~\ref{eq.pertshift}.}

Figure~\ref{fig.wind_eig_overlap} displays meridional slices of $K(r,\mu)$ and the $\delta\omega$ integrand (Equation~\ref{eq.kernelshift}) for the four $m=11$ modes $_{11}^{11}f$, $_{13}^{11}f$, $_{15}^{11}f$, and $_{19}^{11}f$ that have been detected as OLRs in the rings.
}
The high-$\ell$ f-modes are confined close to the surface, and lower latitudinal wavenumbers $\ell-m$ are confined closer to the equator. Consequently the sectoral modes predominantly sample Saturn's prograde equatorial jet, with tesseral ($\ell-m>0$) modes sampling {rotation both at the prograde equator and alternating prograde/retrograde rotation at higher latitudes} (D21). Furthermore the fact that $K(r,\mu)$ is strictly positive means that prograde regions always increase mode frequency, while retrograde regions decrease mode frequency. The frequency shifts induced by winds are thus typically positive for high-degree f-modes, albeit with smaller magnitudes toward larger $\ell-m$ as a result of offsetting contributions from prograde and retrograde winds. Note also that the equatorial symmetry $K(r,\mu)=K(r,-\mu)$ implies that the small antisymmetric components of the winds have \emph{zero} first order effect on the frequency of any mode thanks to the vanishing integral~\ref{eq.kernelshift}. 

{In our full calculation (Equation~\ref{eq.pertshift}), the term $\mathcal U_{ii}^D$ can be nonzero and is largest for high $\ell$, high $m$ modes, where it nonetheless does not exceed magnitudes $\sim10^{-6}-10^{-5}~\omd$ and can safely be ignored.
A complication would arise should the asymmetric wind cause an avoided crossing between a mode of interest and another mode with opposite equatorial parity.}
As shown by D21, antisymmetric components of the wind can in fact mix low-degree f-modes with nearby low-degree g-modes with opposite parity, a phenomenon {that does not appear to play a role for any of the modes that we focus on here}.

Figure~\ref{fig.wind_eig_overlap_1d} presents a one-dimensional view of these ideas for the same three modes, emphasizing the latitudinal overlap between these modes' rotation kernels and the cloud-level wind speed.

\subsection{Parameter estimation}\label{sec.methods.mcmc}
We proceed much as in MF21, calculating the total likelihood as the product $L=L_{\rm seis}\times L_{\rm grav}$ of two multivariate Gaussians such that
\begin{equation}
    \label{eq.likelihood}
    \ln L = 
    - \sum_i \frac{(\omp^{\rm calc} - \omp^{\rm obs})^2}{2\,\Delta\omp^2} 
    - \sum_{n=1}^3 \frac{(J_{2n}^{\rm calc} - J_{2n}^{\rm obs})^2}{2\,\Delta J_{2n}^2}.
\end{equation}
{
$\Delta\omp$ is taken to be uniform at $0.3\ \degd$, approximately $0.02\%$ of the pattern speeds themselves, slightly larger than the fiducial observational uncertainty $\approx0.1\ \degd$ (Section~\ref{sec.methods.modes}). The slightly larger uncertainty aids convergence of the sampling algorithm by retaining a larger fraction of proposed models. We also compare against an alternative sample with an inflated uncertainty $\Delta\omp=3\ \degd$ ($0.2\%$), and a sample that neglects the ring seismology altogether, to emphasize the value of precision seismology for constraining the rotation.
The adopted gravity uncertainties $\Delta J_{2n}$ are given in Section~\ref{sec.methods.rigidmodel}.
}

We consider only uniform distributions for prior probability, so the posterior probability is simply proportional to $L$ inside the allowed prior volume and zero outside of it. 
Aiming to consider the largest plausible prior volume, we adopt the following:
\begin{enumerate}
    \item $\zout\in[0.015,0.3]$, i.e., the outer envelope metallicity is at least solar. Since $Z$ is assumed to increase with depth, $\zout$ is inevitably bounded from above by the mean metallicity consistent with Saturn's mean density, approximately $\langle Z\rangle\lesssim0.2$  \citep[e.g.,][]{1999P&SS...47.1183G} or very conservatively 0.3.
    \item $\rout\in[0,1]$; in practice $J_{2n}$ and the $m=2$ ring seismology constraint are expected to strongly favor $\rout\sim0.6$ (MF21).
    \item $\yin\in[0.275, 1]$; the helium distribution lies between the limiting cases of homogeneous at the protosolar He/H fraction ($\yin=\yout=0.275$) and maximally stratified ($\yin=1$; $\yout<0.275$).
    \item $2\pi/\Omega_{\rm S}\,({\rm min})\in[630,640]$, spanning the range of estimates for Saturn's deep spin from various sources \citep{1981GeoRL...8..253D,2009Natur.460..608R,2015Natur.520..202H,2019ApJ...879...78M}. This rules out the longer periods measured in Saturn's kilometric radiation \citep[e.g.,][]{2006Natur.441...62G} as being directly tied to Saturn's deep rotation; our results will validate this choice.
    \item $d/R_{\rm eq}\in[0.05,0.2]$. The lower limit is nonzero to ensure that our radial decay function does not artificially modify the cloud-level winds. Models near this lower limit have f-mode spectra very similar to the the rigidly rotating case (Figure~\ref{fig.audit_rigid}), {suffering a clear systematic mismatch in the sectoral $(\ell=m)$ modes and precluding any viable solution} with $d/R_{\rm eq}<0.05$. The maximum decay centroid depth {is chosen to well encompass} the values $\approx0.12\,\req\approx7,000\,{\rm km}$ estimated from analysis of zonal gravity \citep{2019Sci...364.2965I,2019GeoRL..46..616G} and magnetic field strength \citep{2021MNRAS.501.2352G}. This depth is approximately where the electrical conductivity, increasing steeply with depth, becomes sufficiently large \citep{2008Icar..196..653L} that Lorentz forces eliminate any differential flows with velocities comparable the cloud-level winds \citep{2017Icar..296...59C,2018Natur.555..227G}.
\end{enumerate}
We use \verb+emcee+ \citep{2013PSP..125..306F} version 2.2.1 to sample these five parameters against the likelihood of Equation~\ref{eq.likelihood} given the first three even zonal gravity harmonics $J_2-J_6$ from \cite{2019Sci...364.2965I}, normalized to $R_{\rm eq}=60,268$ km, and the pattern speeds of the same 14 patterns that are fit by the rigid model in Figure~\ref{fig.audit_rigid}. 
{These patterns and the additional mode-wave associations used to compare models to observations post-hoc are listed in Table~\ref{tab.associations}.}
{Although the odd gravity harmonics ($J_3$, $J_5$, $\ldots$) and higher order even harmonics ($J_{8}$, $J_{10}$, $\ldots$) dominated by the winds are readily available for all models, these are ignored in the fitting procedure: our wind model only includes a single parameter (decay depth) and hence lacks the flexibility to modulate these harmonics independently from one another (e.g., to modify $J_8$ without sacrificing the fit to $J_2-J_6$). Instead, models are directly constrained by just the first three even harmonics, the remainder compared to after the fact.}

Additional samples introduce a parameter $\fice\in[0,1]$ controlling the mass ratio of ices to silicates (see Section~\ref{sec.methods.rigidmodel}){, but we find that silicate-rich models $f_{\rm ice}\ll1$ are heavily disfavored, significantly overestimating the pattern speed of the $_3^3f$ mode by several $\degd$ and underestimating $|J_4|$ by typically 2 ppm.
Hence in what follows we focus on the successful water ice models with $\fice=1$. 
}
{When discussing distributions (e.g., of model parameters), all numerical ranges are given in terms of their 5\% and 95\% quantiles.}

\section{Results} \label{sec.results}
\subsection{The cloud level wind profile}
\label{sec.results.modifiedwind}
{Joint analysis of gravity and seismology strongly favor the Galanti modified cloud-level winds (Section~\ref{sec.methods.wind}) over the mean profile from \cite{2011Icar..215...62G}.
First, ignoring ring seismology entirely and fitting just $J_2-J_6$, a sample strictly adopting the García-Melendo profile tends to systematically overestimate $J_6$. This overestimate drives the most successful models to minimize the wind contribution $\delta J_6$ by adopting the shallowest wind depths allowed by our prior $\depth\geq0.05$. This can lead to plausible fits to $J_6$, {but poor fits to the higher degree harmonics $J_8$ and $J_{10}$}, and post hoc comparisons to the seismology find poor fits that are akin to the rigid model in Figure~\ref{fig.audit_rigid} at best.}

\begin{figure}
    \centering
    \includegraphics[width=\columnwidth]{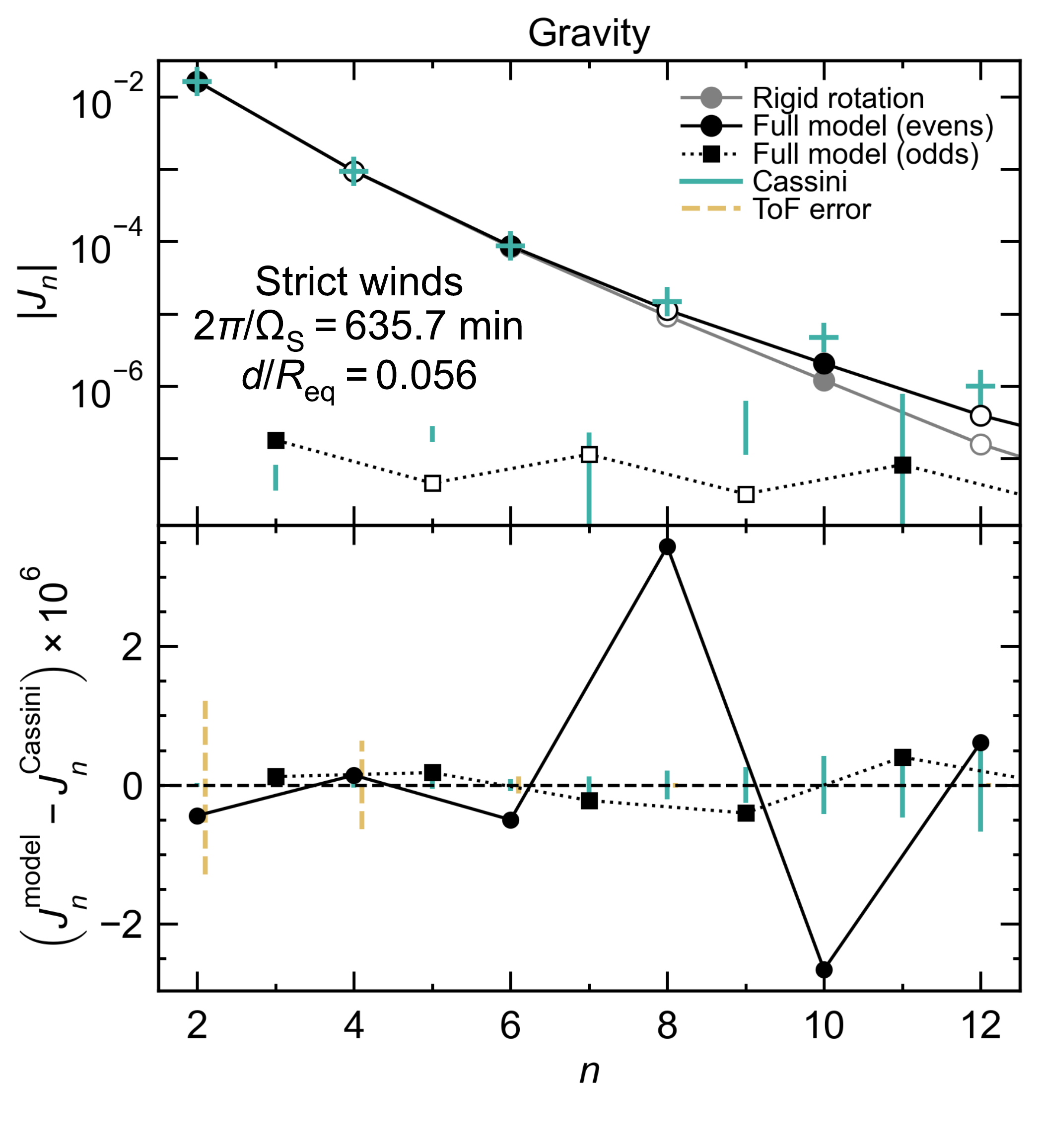}
    \caption{
    {Gravity harmonics $J_n$ (top) and their residuals (bottom) for the best model that strictly satisfies the \cite{2011Icar..215...62G} cloud level wind profile, and is fit to $J_2-J_6$ and the ring seismology.
    Dashed and solid vertical lines represent uncertainty in the ToF calculation and the Cassini measurement uncertainty respectively (see legend). The $J_{2n}$ are overplotted with plus signs.
    This model requires shallow ($d\approx3,400\ {\rm km}$) winds to even marginally fit $J_6$. 
    The poor fit to the odd harmonics, even harmonics $J_{6+}$, and seismic constraints (not pictured) all strongly favor a model with somewhat different surface winds (see text).
    }}
    \label{fig.audit_jn_strict}
\end{figure}

{Repeating this exercise but folding in the seismic constraints yield models that compromise the gravity fit further. 
The gravity spectrum of the best such model is shown in Figure~\ref{fig.audit_jn_strict}, where
the fit to the higher-degree even harmonics $J_6$, $J_8$ and $J_{10}$ is especially poor.
Hence our results support what \cite{2019GeoRL..46..616G} and \cite{2021MNRAS.501.2352G} found by applying a similar functional form for the wind decay: the gravity favor a somewhat different latitudinal profile than the García-Melendo observations. The seismology strengthens this finding.}

{The constancy of Saturn's cloud level winds over multiple observations and long time baselines \citep{2011Icar..215...62G} suggests that this discrepancy arises not from some observational bias but from the simplified decay profiles, including our Equation~\ref{eq.decay}, that have so far been used to interpret the gravity.
}
{The slowly rotating features in Saturnian gravity uncovered by \cite{2022PSJ.....3...61H} also point toward stronger retrograde rotation at depth than the surface winds alone would suggest.}

{
In what follows we limit our attention to the much more successful models that assume the slightly modified ``Galanti reconstructed'' cloud level wind profile depicted in Figure~\ref{fig.wind_eig_overlap_1d}.
As we have mentioned, the principal difference from observations is in the more pronounced retrograde rotation near $|\mu|=0.4-0.6$ ($30\textrm{--}45^\circ$ planetographic latitude), which predominantly affects the frequencies of modes with $\ell-m\geq2$ (Figures~\ref{fig.wind_eig_overlap}-\ref{fig.wind_eig_overlap_1d}).
The modified profile also has a more subtly exaggerated prograde rotation in the next zone at $|\mu|=0.65-0.75$ ($45\textrm{--}55^\circ$), which the same figures show is probed most strongly by modes with $\ell-m\geq4$.
The wind structure at higher latitudes does not greatly affect the spectrum thanks to the limited number of detections with large latitudinal wavenumbers, a point that we return to in this paper's Discussion.
}

\begin{figure*}
\begin{center}
\includegraphics[width=\textwidth]{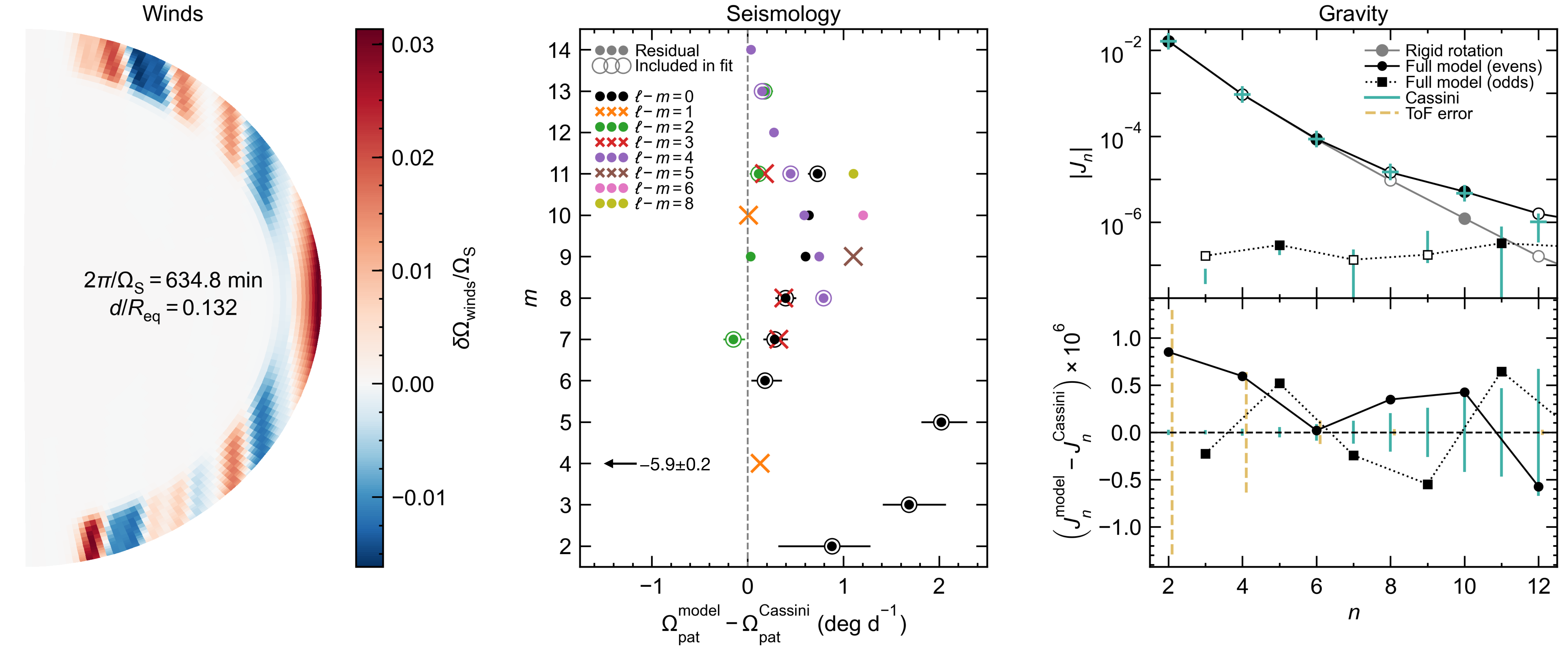}
\caption{
\label{fig.audit_best} 
Wind model (left), pattern speed residuals for {waves} observed at Lindblad {or vertical} resonances {in the C ring} (center), and gravity harmonics (top right) and residuals (bottom right) for our best fitting Saturn model. Different modes are distinguished by their azimuthal wavenumber $m$ (vertical axis) and latitudinal wavenumber $\ell-m$ (see colors {and markers} in legend). 
The circled modes are the high-degree modes calculated on the fly and fit to data during MCMC sampling, their error bars showing the 5-95\% quantiles of residuals obtained by all models across the sample of tens of thousands.
}
\end{center}
\end{figure*}

\subsection{The seismic imprint of Saturn's winds}\label{sec.results.main}
Deep zonal winds bring the f-mode frequencies into substantially better accord with observations. Figure~\ref{fig.audit_best} summarizes the best model's zonal wind structure and displays pattern speed residuals (calculated minus observed) for Saturn-associated spiral density {and bending} waves in the C ring.
The winds ameliorate the worst systematic issues identified in the rigid model, mainly the strong tendency to underestimate the frequencies of sectoral ($\ell=m$) modes relative to higher latitudinal wavenumbers $\ell-m>0$, and the tendency for this bias to get worse toward higher angular degrees (the steep slope of blue points in Figure~\ref{fig.audit_rigid}).
{For the high-degree ($\ell=m\geq6$) modes,} the differentially rotating model's pattern speed residuals span $-0.2$ to $+1.2\ \degd$, a marked improvement from the rigid model's span of $-2$ to $+3\ \degd$ in Figure~\ref{fig.audit_rigid}. This model achieves this with a wind decay depth $\depth=0.132$ and deep spin period $\psat=634.8$ min, the deep spin some 2 min slower than the rigid model as necessary to compensate for the tendency of the predominantly prograde deep winds to shift all modes toward higher frequency. 

{The 6 antisymmetric modes with $\ell-m=1,3,5$ detected via bending waves show a similar quality of fit{, t}heir residuals ranging from 0 to $+1.1\ \degd${.} This improvement over the best rigid model's $+0.2$ to $+1.9$ span in the $\ell-m=1,3,5$ residuals (Figure~\ref{fig.audit_rigid}) is not a result the N-S asymmetric component of Saturn's winds, which \text{have a negligible} effect on the frequency of isolated f-modes (see Section~\ref{sec.methods.dom}). Rather, the improvement is a direct result of the symmetric component of the winds, as well as an indirect result of the same symmetric winds' impact on the $J_{2n}$ and $\ell=0,2,4$ modes that these models optimize.}

{The situation is different for the low-degree ($\ell=m<6$) modes, whose residuals are very similar to the rigidly rotating case (Figure~\ref{fig.audit_rigid}), reflecting these deep-seated modes' relatively low sensitivity to the zonal winds and greater sensitivity to the structure of the deep interior.
Their outstanding $1-6\ \degd$ residuals are much larger than the fiducial pattern speed uncertainty $0.1\ \degd$, likely a result of the composition structure posited here not being perfectly accurate, as discussed in MF21; we revisit this in Section~\ref{sec.discussion}.}
On this topic we also note that the residuals are overwhelmingly positive, a fact that follows in part from the simultaneous fitting of the $J_{2n}$ and in part from the outlying $_4^4f$ mode. Repeating this sample but neglecting the less well understood $_3^3f$, $_4^4f$, and $_5^5f$ modes yields consistent deep spin and wind decay depths, {and does little to alleviate the systematic trends seen in the residuals at higher $m$ (see Appendix~\ref{app.omitm345})}.

\begin{figure}
\begin{center}
\includegraphics[width=\columnwidth]{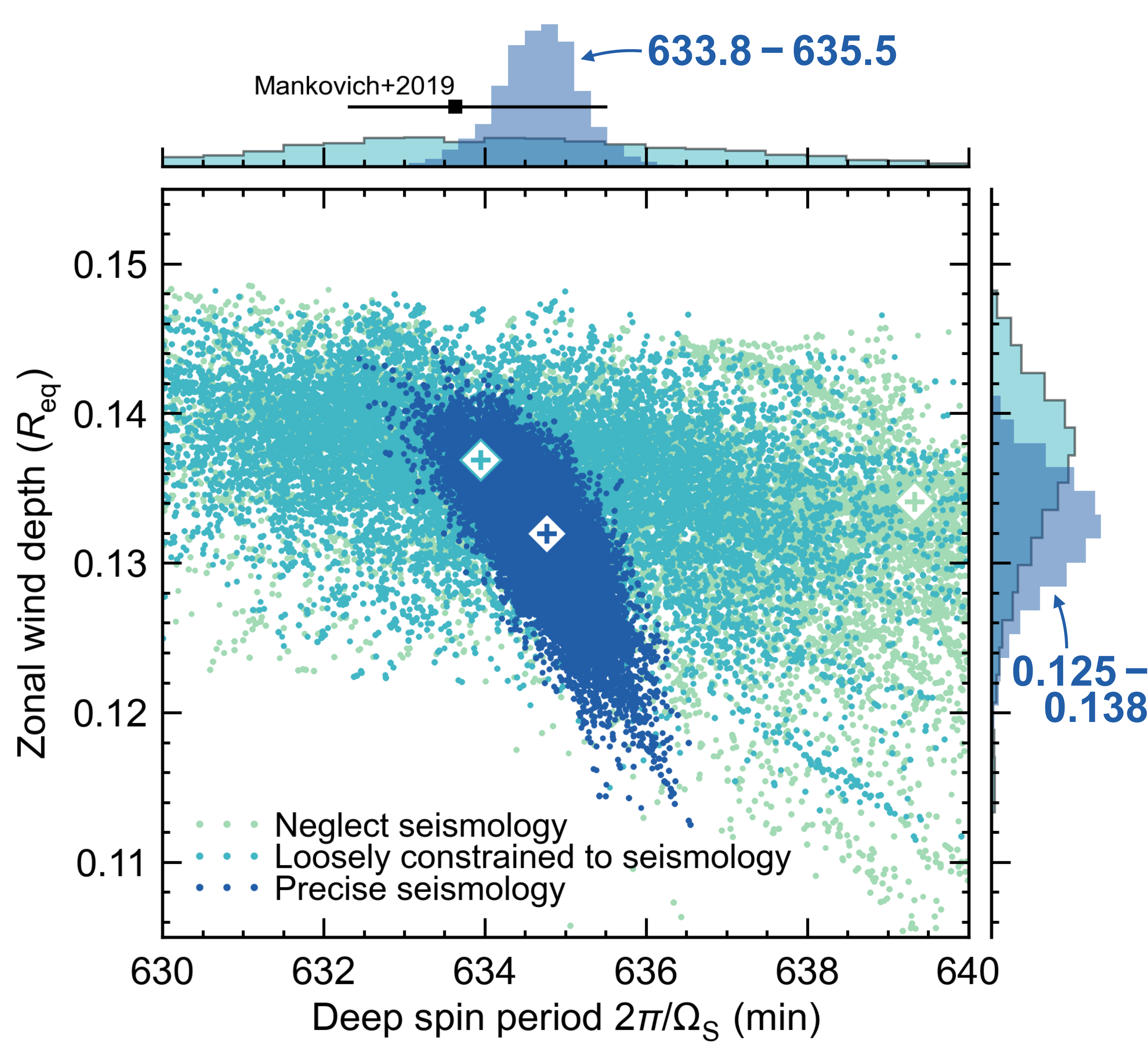}
\caption{
\label{fig.psat-depth} 
{Decay depth of the zonal winds as a function of Saturn's deep spin period. The light blue distribution fits the ring seismology with an exaggerated pattern speed uncertainty $\Delta\Omega_{p,i}=3$ deg/d while the dark blue distribution uses $\Delta\Omega_{p,i}=0.3$ deg/d, slightly larger than the observational uncertainty (see Section~\ref{sec.methods.modes}).
{Plus signs designate the most likely single model from each sample.}
Histograms are marginalized posterior probability distributions.
The faint green distribution in the background ignores the seismology; histograms are omitted for this case.
The error bar laid over the histograms (top) gives the earlier M19 estimate for Saturn's spin period based on rigid models and fewer data.
}
}
\end{center}
\end{figure}

{
Figure~\ref{fig.psat-depth} summarizes the posterior distribution of models in deep spin -- wind decay depth space. Here we compare three samples that all fit $J_2\textrm{--}J_6$ but incorporate progressively more information: the first neglects ring seismology, the second fits ring seismology but with inflated pattern speed uncertainties $\Delta\omp=3\ \degd$, and the final favored sample fits the ring seismology with $\Delta\omp=0.3\ \degd$, a conservative value for the observational uncertainty (see Section~\ref{sec.methods.modes}). 
All three distributions occupy a similar range in $\depth$, reflecting the fact that the decay depth is most strongly constrained by $J_6$. 
However, the $\depth$ distribution is both systematically lower and modestly better constrained in the favored (``Precise seismology'') sample, demonstrating that the ring pattern speeds also constrain the depth of the differential rotation. 
For the favored sample we find a median depth $\depth=0.132$ (consistent with the maximum likelihood estimate appearing in Figure~\ref{fig.audit_best}) and width $\depth=0.125\textrm{--}0.138$.
}

{
The other main result in Figure~\ref{fig.psat-depth} is that compared to models indifferent to or weakly informed by the ring seismology, the favored sample has a much better constrained deep spin period. 
From this sample we find a median deep spin $\psat=634.7$ min (close to the maximum-likelihood $634.8$ min period of Figure~\ref{fig.audit_best}) and width $633.8\textrm{--}635.5$ min.
In the case of the more loosely constrained sample, the distribution peaks at similar deep spins but is unhelpfully broadly distributed, emphasizing the critical importance of the frequency precision afforded by Cassini ring seismology.
The final sample that ignores ring seismology is nearly uniformly distributed in $\psat$, although the paucity of constraints in this case (3 $J_{2n}$ compared to 5 free parameters) lead to a sample that is not statistically well converged even after $\sim10,000$ MCMC iterations. For this reason we refrain from plotting histograms for this case, but it is noteworthy that high likelihood models are found throughout the allowed range of $\psat$. 
Neglecting seismology, the single most likely model of the pure-gravity sample attains a slow deep spin $\psat=639.3$~min. This underscores the value of ring seismology for eliminating much of this parameter space for Saturn's deep rotation, including periods $\approx10^{\rm h}39^{\rm m}$ and longer that appear in Saturn's kilometric radiation \citep[e.g.,][]{1981GeoRL...8..253D,2006Natur.441...62G,2018GeoRL..45.7297Y}.
}

\begin{figure*}
\begin{center}
\includegraphics[width=\textwidth]{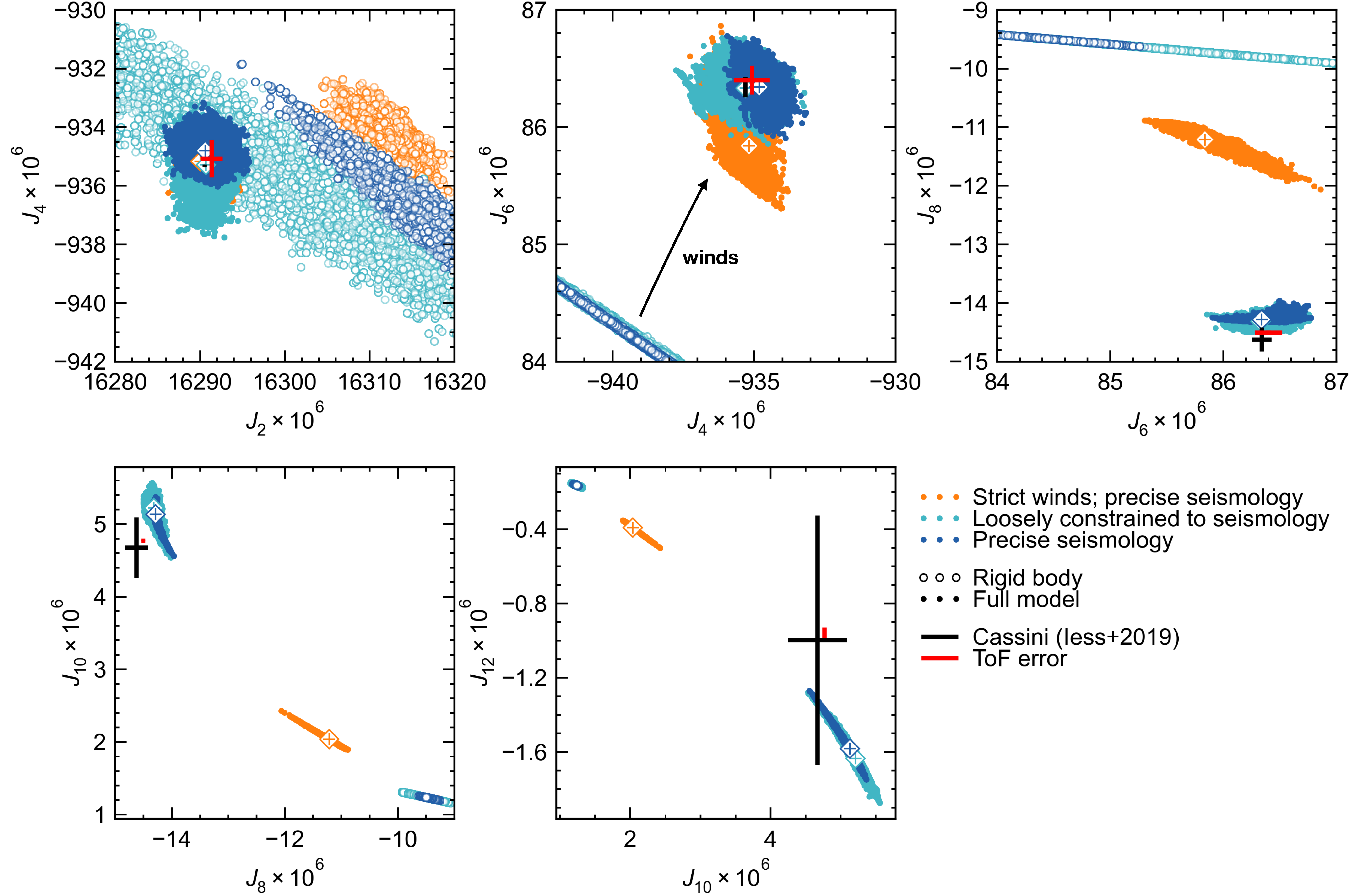}
\caption{
\label{fig.j2n_scatter} 
{Even gravity harmonics through $J_{12}$ for the three samples indicated in the legend. Open {circles} are rigid body components and filled {circles} are total harmonics (rigid plus wind contributions). ``Strict winds'' refers to the \cite{2011Icar..215...62G} cloud level velocity profile; the other two samples assume the Galanti reconstructed profile (see Figure~\ref{fig.wind_eig_overlap_1d} and Section~\ref{sec.methods.wind}). Diamonds with plus signs show the best model from each sample. {Black crosses show the \cite{2019Sci...364.2965I} $3\sigma$ measurement uncertainty and the red crosses show the ToF7 uncertainty (Section~\ref{sec.methods.rigidmodel}), slightly offset here for readability. For ease of comparison, this figure normalizes harmonics to the conventional reference radius 60,330 km.}
}
}
\end{center}
\end{figure*}

{
Figure~\ref{fig.j2n_scatter} shows the full distribution of the $J_{2n}$ in three of our samples, separating the contributions from the rigid background and the differential rotation. Here it is clear that even across the entire parameter space available to the models, models with strict cloud-level winds struggle to fit $J_6$ (by going to very shallow winds; see Section~\ref{sec.results.modifiedwind}) and failing more acutely to fit $J_{8+}$.
}

\begin{figure}
\begin{center}
\includegraphics[width=\columnwidth]{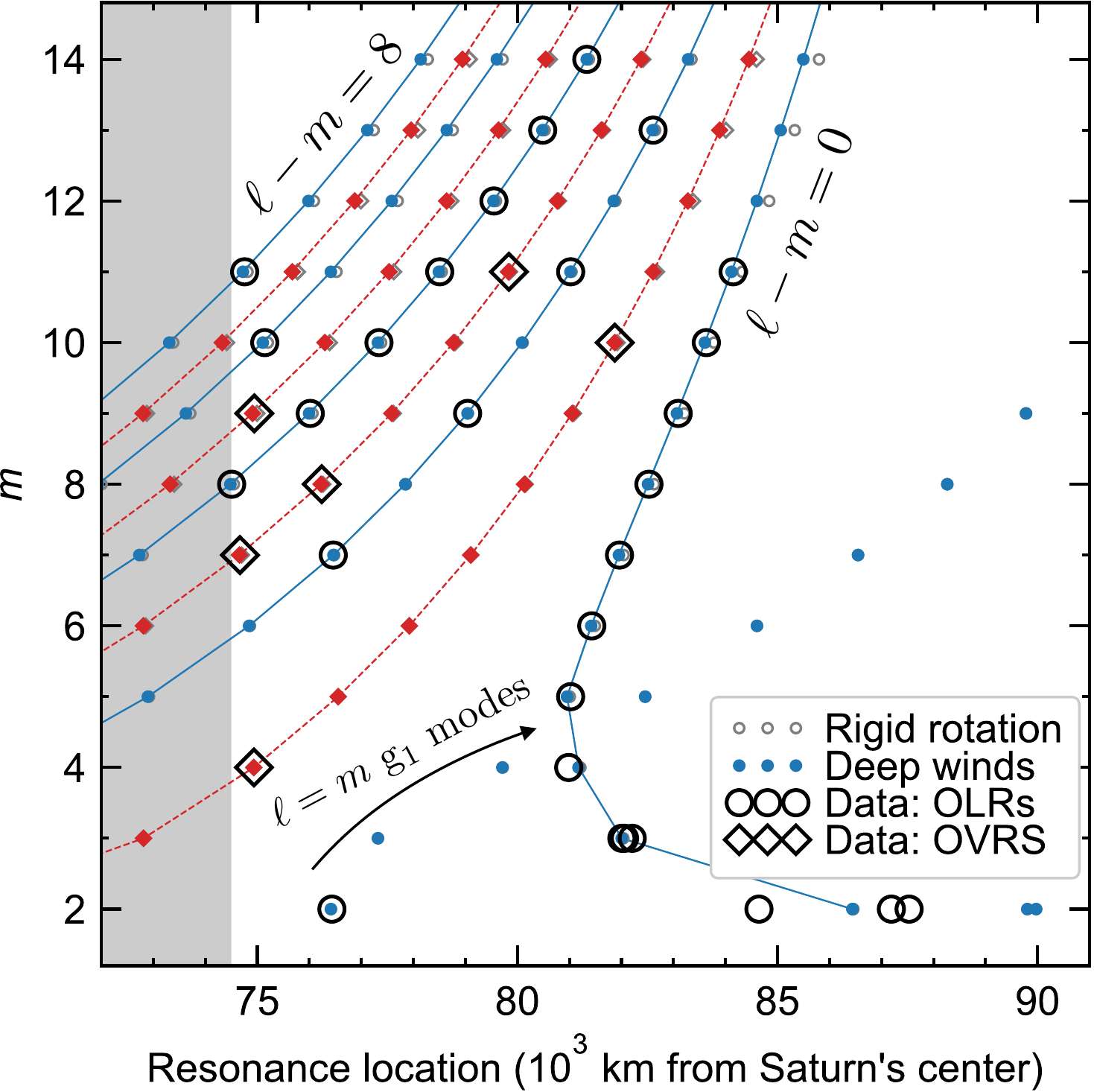}
\caption{\label{fig.m-rl} Resonance locations predicted for the best model with winds (small filled circles) and for the same model but ignoring winds (small open circles), compared to all observed C ring outer Lindblad resonances (open black circles) and outer vertical resonances (open black diamonds) excited by Saturn modes (\citealt{2021Icar..37014660F} and references therein).
Solid lines connect modes with even $\ell-m$ capable of producing density waves at Lindblad resonances; dashed lines connect modes with odd $\ell-m$ capable of producing bending waves at vertical resonances.
The shaded region is the D ring where resonances are not readily observable.
Rigid and differential rotation are indistinguishable at this scale for the $\ell=m$ g$_1$ modes and $m\lesssim4$ modes.
}
\end{center}
\end{figure}

A summary of the best model's fit to observations, in terms of absolute resonance locations in Saturn's rings, is displayed in Figure~\ref{fig.m-rl}. Here all model f-modes with $m=5-14$ and $\ell-m\leq8$ are shown; for $m=2-4$ the distinction between f- and g-modes becomes less clear (see MF21, D21) and we plot only modes with strong surface gravitational potential perturbations, with lines connecting the modes with the strongest perturbation in each $m$. Among these $m=2-4$ modes only {a subset of detections are} used to constrain the model, and the remainder of these low-$m$ modes do not represent an exhaustive search, which requires a great deal of radial and spectral resolution (see Section~\ref{sec.methods.modes}). Ongoing work will revisit these in detail. Toward higher $m$, this broad-scale view shows the overall success of the model; in particular, the wind model brings the sectoral ($\ell=m$) modes into excellent agreement with observations.

\begin{figure*}
\begin{center}
\includegraphics[width=\textwidth]{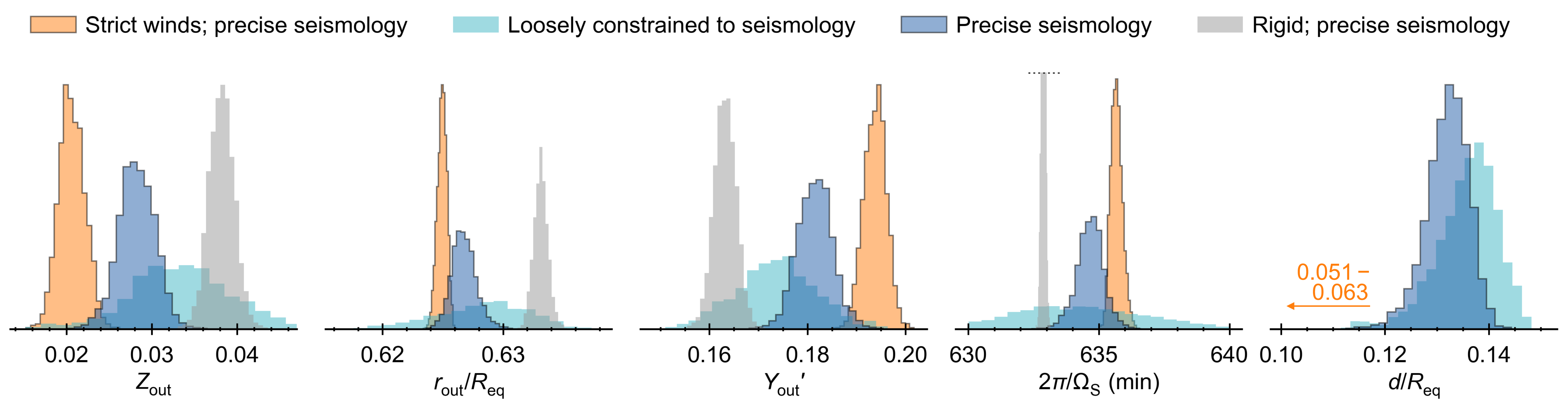}
\caption{\label{fig.1dpost} Posterior probability distributions of model parameters, {normalized to equal area}, compared {among the samples appearing in Figure~\ref{fig.j2n_scatter}}. The grey distribution shows the rigidly rotating sample corresponding to Figure~\ref{fig.audit_rigid}{; its narrow $\psat$ distribution is cut off somewhat for readability of the other distributions}.
{The favored (dark blue) sample is the one} represented in Figures~\ref{fig.audit_best}$-$\ref{fig.m-rl}.
}
\end{center}
\end{figure*}

\subsection{Composition}
\label{sec.results.composition}
Figure~\ref{fig.1dpost} reports marginalized posterior distributions for 
{the same three samples as in Figure~\ref{fig.j2n_scatter}, plus the rigidly rotating sample.}
In all cases, zonal gravity constrains these models to 1-3 times {proto}solar metallicity {(here $Z=0.014$; e.g., \citealt{2009ARA&A..47..481A})} in the envelope, at odds with the $9\times$ solar carbon enrichment from spectroscopy \citep{2009Icar..199..351F}. Higher envelope metallicities lead to unrealistically large magnitudes of the $J_{2n}$.
Part of this discrepancy may be attributable to uncertainties in the hydrogen EOS in the vicinity of metallization, where fractional density errors of order a few percent are plausible and can lead to substantial increases in the envelope metallicity \citep{BM22}. 
{Higher envelope entropy (parameterized in adiabatic models as an artificially high 1-bar temperature) also tends to allow more realistic metallicities \citep{2021PSJ.....2..241N,2022A&A...662A..18M} by increasing the mean molecular weight necessary to achieve the density required by the $J_{2n}$. The source of this tension may ultimately be a combination of errors in the hydrogen EOS and some degree of superadiabaticity in the outer envelope.}

As in MF21 we find no solutions with uniform $Y^\prime$: in all models the zonal gravity, particularly $J_4$, favors a moderately helium-depleted envelope $\yout=0.15-0.20$, consistent with the measurement by \cite{2018Icar..307..161K} combining Cassini UVIS and CIRS data but not the stronger depletion implied by \cite{2020PSJ.....1...30A}. Models with $\yout\gtrsim0.20$ are generally compatible with gravity but would require sub-solar $\zout$ and lie outside our prior volume (see Section~\ref{sec.methods.mcmc}). Models with a stronger helium stratification $\yout<0.14$ accommodate more strongly super-solar envelope heavy element enrichments, but acquire unrealistically high $|J_4|$. In light of Saturn's $9\times$ solar carbon enrichment, it may be that refinements in the EOS will lead to a reduced density for hydrogen at depth, allowing for a more metal-rich and helium-poor outer envelope than the models have achieved to date. 

\subsection{Moment of inertia}
\label{sec.results.moi}
{Saturn's moment of inertia is of particular interest from the perspective of a proposed resonance between Saturn's spin axis precession and Neptune's orbital precession, a commensurability that could help to explain Saturn's high obliquity \citep{2004AJ....128.2501W,Saillenfest2021} and the recent formation of the rings \citep{doi:10.1126/science.abn1234}. Our differentially rotating models have a narrow range of polar moments of inertia $C/MR_{\rm eq}^2=0.2188-0.2191$, consistent with the upper end of estimates from the gravity modeling from \cite{doi:10.1126/science.abn1234}, and approximately $0.5\%$ smaller than the value required for the resonance to be active today.}
{This supports the notion that the resonance is inactive today, but may have been active in the past.
The significantly higher values estimated by \cite{2020ApJ...891..109M} may be attributable in part to that work's assumption that the magnitude of the $\delta J_{2n}^{\rm winds}$ do not exceed the \cite{2017ApJ...843L..25G} predictions in advance of the Cassini Grand Finale. 
The now compelling evidence for deep zonal winds and their considerable perturbations to zonal gravity (see Figure~\ref{fig.tgwe_oblate_djn_winds}) warrants a revisiting of the probabilistic Saturn interior modeling advanced by \cite{2020ApJ...891..109M}.}

\section{Discussion} \label{sec.discussion}
{
Under our assumed functional form (Equation~\ref{eq.decay}) for the wind decay, we find that a somewhat modified profile for the surface winds was required for satisfactory solutions to the gravity field, a requirement made even more acute when fitting gravity and seismology jointly (Section~\ref{sec.results.modifiedwind}).
Meanwhile, it appears that the observed winds \citep{2011Icar..215...62G} are robust across imaging data sets and observational epochs.
Other studies have addressed similar tension by appealing to latitudinally dependent wind depths, for instance in the Jupiter models of \cite{BM22}.
We suggest that Equation~\ref{eq.decay} is failing to capture some nontrivial spatial dependence below the cloud level that tends to exaggerate the retrograde rotation of the latitudes just off the main prograde jet, such as a deeper penetrating wind localized to those latitudes.
}
{Future work exploring these possibilities will benefit from directly fitting the full set of even zonal harmonics including $J_{8+}$.}

{
Of Saturn's odd zonal gravity harmonics, at least $J_3$, $J_5$, and $J_9$ are statistically different from zero \citep{2019Sci...364.2965I,2022AJ....164..199J}, a signal arising entirely from the antisymmetric component of the differential rotation. 
We have not fit these harmonics directly in our calculations because our simplified wind model (Equation~\ref{eq.decay}) lacks the flexibility required to alter these greatly, and more importantly the odd component of the winds has {a negligible} effect on the frequency of isolated Saturn modes (Section~\ref{sec.methods.dom}), {meaning that gravity remains the best probe of N-S asymmetry in the differential rotation}.
Nonetheless we find a plausible fit to the odd $J_n$ (Figure~\ref{fig.audit_best}) when using the Galanti reconstructed winds (Section~\ref{sec.results.modifiedwind}). This is expected since that profile was allowed to deviate from the García-Melendo profile slightly in order to match the $J_n$.
We note that the strict winds yield a much worse fit to the higher degree harmonics (Figure~\ref{fig.audit_jn_strict}), confirming the finding from \cite{2019GeoRL..46..616G} and \cite{2021MNRAS.501.2352G} that the mean cloud level wind profile from \cite{2011Icar..215...62G} and a simple decay law like Equation~\ref{eq.decay} are not a complete description for the flow within Saturn. Nonetheless the joint analysis of seismology and gravity clearly strengthens the case for deep winds in Saturn, and the seismology in particular is an effective anchor for the deep spin {period} (Figure~\ref{fig.psat-depth}).
}

{New estimates for Saturn's gravity moments obtained by \cite{2022AJ....164..199J} depart from the values reported by \cite{2019Sci...364.2965I}, in many cases by an amount greater than the formal statistical $3\sigma$ uncertainty associated with each measurement. While the deviations for $J_2$, $J_4$ and $J_6$ are too small to fundamentally alter our results here, they do suggest that the $J_n$ measurements can be affected by systematic uncertainties that we hope may be better quantified in the future. In light of this, we encourage modeling efforts in the meantime to allow for the broadest plausible range of gravity moments.
}

{
Our estimate for the deep spin {period} $\psat=633.8-635.5\ {\rm min}$ is consistent with estimates from atmospheric Rossby waves \citep{2009Natur.460..608R} and Saturn's gravity field and oblateness \citep{2019ApJ...879...78M}.
Our range of 1.7 min in retrieved spin periods is approximately twice as precise as the 3.2 min estimate from M19. 
The improvement follows from the application of more precise and accurate methods for computing the $J_{2n}$ and the additional constraint from $J_6$ (Section~\ref{sec.methods.rigidmodel}$-$\ref{sec.methods.djn}), as well as more precise methods for computing mode frequencies and the many additional modes frequencies used to constrain the model (Section~\ref{sec.methods.modes}).
That the improvement was so modest at a factor of just $\approx2$ follows from the additional degeneracies that our model encounters, allowing 5 degrees of freedom compared to M19's 2 to 3.
Our precision is also limited in part by the broad tolerances we must allow for $J_{2n}$ owing to the imperfect accuracy of the theory of figures (see Section~\ref{sec.methods.rigidmodel} and Figure~\ref{fig.audit_best}, bottom right panel). It may be worthwhile to bring more accurate treatments of the rigid body component of the gravity field, such as the concentric Maclaurin spheroid method \citep{2013ApJ...768...43H}, to bear on this line of seismological inference.
}

{
Even more critical are the statistically significant frequency residuals we find even in our most successful models (Figure~\ref{fig.audit_best}). 
These residuals are worst at low degree $\ell<6$, where Saturn's less certain deep interior structure plays a large role (MF21, D21).
Here aspects of the spectrum (e.g., the $m=2$ spectrum; see MF21) may be better explained by distinct g-mode cavities arising from disconnected gradients in heavy element and helium abundance respectively. More work is required to explore the consequences of such a structure on the other $\ell<6$ modes; we return to this below.
In our model the worst fitting pattern speed, that of the mode $_4^4f$, is influenced by an avoided crossing with the mode $_4^4g_1$ (MF21, D21, Figure~\ref{fig.m-rl}), effectively causing a frequency repulsion between the two \citep{2014Icar..242..283F}. 
Resolving this mismatch would require a weaker interaction between the two, and this may be a useful clue for future work on the deep interior structure.
}

{
Our best pattern speed residuals are also significant at $\ell\geq6$, where modes are confined to the envelope, indicating outstanding issues with the thermal, composition, or rotation profile in the outer regions ($r>\rout\sim0.6$) that are challenging to resolve given the inherent degeneracy of these different factors. This situation is reminiscent of the persistent difficulty encountered in fitting Jupiter and Saturn interior models to gravity while retaining the super-solar envelope heavy element enrichments indicated by spectroscopic measurements \citep{2022arXiv220504100G}. 
Ad hoc resolutions have been framed in terms of either (1) unrealistically high 1-bar temperatures, a simple means of parameterizing some larger interior entropy than adiabats satisfying these planet's 1-bar temperatures would predict; or (2) a systematic error in the density of hydrogen expected from high-pressure physics, especially at pressures 0.1 to 1 Mbar \citep{2021PSJ.....2..241N,2022A&A...662A..18M,BM22}.
Jupiter models proposed by \cite{2019ApJ...872..100D} jettison these possibilities in favor of a more physically consistent, albeit intricate, interpretation that appeals to composition and entropy gradients.
Even with the availability of precision Juno and Cassini gravity the overall picture remains unclear, but composition gradients and nonadiabaticity will certainly need to be considered as the interpretation of the seismology matures.
}

{
The ring seismology carries more information about low latitudes than high latitudes, a consequence of the fact that modes with latitudinal wavenumbers $\ell-m=0,2,4$ are apparently more readily detected than higher wavenumbers. This is partly a geometric effect: the torques exerted by a resonance decrease with $\ell-m$ (M19). Preferential excitation of low wavenumbers may also play a role.
\cite{2021Icar..37014660F} uncovered the highest latitudinal wavenumber mode to date in $_{19}^{11}f$ (W74.76).
We note that this mode fits well for the rigid model in Figure~\ref{fig.audit_rigid} but is somewhat overestimated in our favored model including winds (Figure~\ref{fig.audit_best}).
This indicates that in reality this mode must suffer more self-cancellation in the $\delta\omega$ integrand (Equation~\ref{eq.pertshift}) than in our wind model, perhaps suggesting that the prograde rotation at $|\mu|=0.65-0.85$ is exaggerated in the ``Galanti reconstructed'' profile.
Future work that incorporates more freedom in the wind model will be able to use this and all the available constraints spanning $m$ and $\ell-m$, and take full advantage of the high-degree even ($J_{8+}$; Figure~\ref{fig.j2n_scatter}) and odd ($J_3$, $J_5$, $J_9$; Figure~\ref{fig.audit_best}) gravity harmonics, to refine our understanding of the flow structure.
}

{The present models, like those in MF21, assume for simplicity that the helium and heavy element gradients occupy the same region of the interior. In reality an adiabatic convection zone may separate the stably stratified dilute core from a shallower, stably stratified, helium gradient. Such a convection zone would not necessarily fundamentally alter the $_2^2g_1$ mode that we identify with the wave W76.44, but may aid in interpreting other aspects of the low-degree spectrum, such as the proximity of the $m=2$ pattern W84.64 to the W87.19/Maxwell pair that MF21 (or our present models; see Figure~\ref{fig.m-rl}) could not fully explain. This proximity could be achieved more easily with two distinct g-mode cavities, where for instance an $n=2$ g-mode confined mostly to the diffuse core would interact more weakly (and suffer less frequency repulsion) with the f-mode than in our present model.
This multiple-cavity structure is also attractive from the perspective of Saturn's magnetic field: the intervening convective region could be the seat of Saturn's magnetic dynamo, with an overlying stably stratified helium gradient helping to explain the axisymmetry \citep{1982GApFD..21..113S,2010GeoRL..37.5201S,2021AGUA....200318Y} and lack of secular variability \citep{2021PSJ.....2..181M} of the external field. Future work focusing on the low-degree ring seismology will be able to gauge the viability of such a structure.}

On a different note, a future mission to the ice giants could lead to the discovery of Uranian seismicity via resonances its ring system \citep{2022PSJ.....3..194A}, where even a single mode detection would offer a precious anchor for Uranus' poorly constrained interior rotation and stratification \citep{2010Icar..210..446H,2022PSJ.....3...88M}. The same can be said for Neptune oscillations, which may be detectable in Neptune's tenuous ring system, albeit a more remote possibility. We note that in comparison to Saturn, the ice giants' slower fractional rotation $\Omega/\omd\lesssim20\%$ implies that frequencies measured in inertial space receive a smaller contribution from the Doppler shift, resulting in a larger degeneracy between rotation and the planet's uncertain interior structure.

\section{Summary} \label{sec.conclusion}
{We have applied oblate, partially stably stratified Saturn interior models to the wealth of gravity and ring seismology data made available by Cassini, to conclude that
\begin{enumerate}
    \item Saturn's cloud-level winds penetrate deeply before decaying rapidly at a depth of $0.125-0.138\ \req$, or $7,530-8,320$ km, with a median depth $0.132\ \req=7,960$ km. These are consistent with estimates from gravity and magnetic field science \citep{2019GeoRL..46..616G,2021MNRAS.501.2352G}.
    \item Minor changes to the \cite{2011Icar..215...62G} cloud-level wind profile appear to be necessary, likely indicating that Saturn's wind decay is more complicated than the functional form we have assumed in Equation~\ref{eq.decay}.
    \item Saturn's rigidly rotating deep interior has a spin period of $633.8-635.5$ min ($10^{\rm h}33^{\rm m}48^{\rm s}-10^{\rm h}35^{\rm m}30^{\rm s}$), or a pattern speed $815.7-818.0\ \degd$.
    The median spin period is 634.7 min ($10^{\rm h}34^{\rm m}48^{\rm s}$) for a pattern speed $816.8\ \degd$.
    The new estimate is consistent with \cite{2019ApJ...871....1M} but more precise thanks to additional ring seismology constraints and improved methods for calculating the gravity and normal modes, both in the rigidly rotating limit and in the presence of differential rotation. 
\end{enumerate}
Despite these gains, the persistent pattern speed residuals seen for our most successful model, at both low and high angular degree (Figure~\ref{fig.audit_best}), show that the Cassini ring seismology data are still not perfectly understood. Future analysis of the same data but allowing for more freedom in the composition, temperature, and rotation profiles should bring Saturn's interior into clearer focus.}

\section*{Acknowledgments}
We thank the anonymous referees for their helpful critiques and Eli Galanti for sharing his Saturn wind solution.
We are grateful for support from the Caltech Center for Comparative Planetary Evolution. 
JWF gratefully acknowledges support from the Sloan Foundation through grant FG-2018-10515, and JWD acknowledges support from the Natural Sciences and Engineering Research Council of Canada (NSERC) [funding reference \#CITA 490888-16].


\begin{thebibliography}{}
    \expandafter\ifx\csname natexlab\endcsname\relax\def\natexlab#1{#1}\fi
    \providecommand{\url}[1]{\href{#1}{#1}}
    \providecommand{\dodoi}[1]{doi:~\href{http://doi.org/#1}{\nolinkurl{#1}}}
    \providecommand{\doeprint}[1]{\href{http://ascl.net/#1}{\nolinkurl{http://ascl.net/#1}}}
    \providecommand{\doarXiv}[1]{\href{https://arxiv.org/abs/#1}{\nolinkurl{https://arxiv.org/abs/#1}}}
    
    \bibitem[{{Achterberg} \& {Flasar}(2020)}]{2020PSJ.....1...30A}
    {Achterberg}, R.~K., \& {Flasar}, F.~M. 2020, \psj, 1, 30,
      \dodoi{10.3847/PSJ/ab9cb6}
    
    \bibitem[{{Aerts} {et~al.}(2010){Aerts}, {Christensen-Dalsgaard}, \&
      {Kurtz}}]{2010aste.book.....A}
    {Aerts}, C., {Christensen-Dalsgaard}, J., \& {Kurtz}, D.~W. 2010,
      {Asteroseismology} (Dordrecht: Springer), \dodoi{10.1007/978-1-4020-5803-5}
    
    \bibitem[{{A'Hearn} {et~al.}(2022){A'Hearn}, {Hedman}, {Mankovich}, {Aramona},
      \& {Marley}}]{2022PSJ.....3..194A}
    {A'Hearn}, J.~A., {Hedman}, M.~M., {Mankovich}, C.~R., {Aramona}, H., \&
      {Marley}, M.~S. 2022, \psj, 3, 194, \dodoi{10.3847/PSJ/ac82bb}
    
    \bibitem[{{Asplund} {et~al.}(2009){Asplund}, {Grevesse}, {Sauval}, \&
      {Scott}}]{2009ARA&A..47..481A}
    {Asplund}, M., {Grevesse}, N., {Sauval}, A.~J., \& {Scott}, P. 2009, \araa, 47,
      481, \dodoi{10.1146/annurev.astro.46.060407.145222}
    
    \bibitem[{{Bailli{\'e}} {et~al.}(2011){Bailli{\'e}}, {Colwell}, {Lissauer},
      {Esposito}, \& {Srem{\v{c}}evi{\'c}}}]{2011Icar..216..292B}
    {Bailli{\'e}}, K., {Colwell}, J.~E., {Lissauer}, J.~J., {Esposito}, L.~W., \&
      {Srem{\v{c}}evi{\'c}}, M. 2011, \icarus, 216, 292,
      \dodoi{10.1016/j.icarus.2011.05.019}
    
    \bibitem[{{Ballot} {et~al.}(2012){Ballot}, {Ligni{\`e}res}, {Prat}, {Reese}, \&
      {Rieutord}}]{2012ASPC..462..389B}
    {Ballot}, J., {Ligni{\`e}res}, F., {Prat}, V., {Reese}, D.~R., \& {Rieutord},
      M. 2012, in Astronomical Society of the Pacific Conference Series, Vol. 462,
      Progress in Solar/Stellar Physics with Helio- and Asteroseismology, ed.
      H.~{Shibahashi}, M.~{Takata}, \& A.~E. {Lynas-Gray}, 389.
    \newblock \doarXiv{1109.6856}
    
    \bibitem[{{Bonazzola} {et~al.}(1998){Bonazzola}, {Gourgoulhon}, \&
      {Marck}}]{1998PhRvD..58j4020B}
    {Bonazzola}, S., {Gourgoulhon}, E., \& {Marck}, J.-A. 1998, \prd, 58, 104020,
      \dodoi{10.1103/PhysRevD.58.104020}
    
    \bibitem[{{Brygoo} {et~al.}(2021){Brygoo}, {Loubeyre}, {Millot}, {Rygg},
      {Celliers}, {Eggert}, {Jeanloz}, \& {Collins}}]{2021Natur.593..517B}
    {Brygoo}, S., {Loubeyre}, P., {Millot}, M., {et~al.} 2021, \nat, 593, 517,
      \dodoi{10.1038/s41586-021-03516-0}
    
    \bibitem[{{Cao} {et~al.}(2020){Cao}, {Dougherty}, {Hunt}, {Provan}, {Cowley},
      {Bunce}, {Kellock}, \& {Stevenson}}]{2020Icar..34413541C}
    {Cao}, H., {Dougherty}, M.~K., {Hunt}, G.~J., {et~al.} 2020, \icarus, 344,
      113541, \dodoi{10.1016/j.icarus.2019.113541}
    
    \bibitem[{{Cao} \& {Stevenson}(2017{\natexlab{a}})}]{2017JGRE..122..686C}
    {Cao}, H., \& {Stevenson}, D.~J. 2017{\natexlab{a}}, Journal of Geophysical
      Research (Planets), 122, 686, \dodoi{10.1002/2017JE005272}
    
    \bibitem[{{Cao} \& {Stevenson}(2017{\natexlab{b}})}]{2017Icar..296...59C}
    ---. 2017{\natexlab{b}}, \icarus, 296, 59, \dodoi{10.1016/j.icarus.2017.05.015}
    
    \bibitem[{{Debras} \& {Chabrier}(2019)}]{2019ApJ...872..100D}
    {Debras}, F., \& {Chabrier}, G. 2019, \apj, 872, 100,
      \dodoi{10.3847/1538-4357/aaff65}
    
    \bibitem[{{Desch} \& {Kaiser}(1981)}]{1981GeoRL...8..253D}
    {Desch}, M.~D., \& {Kaiser}, M.~L. 1981, \grl, 8, 253,
      \dodoi{10.1029/GL008i003p00253}
    
    \bibitem[{{Dewberry} {et~al.}(2022){Dewberry}, {Mankovich}, \& {Fuller}}]{D22}
    {Dewberry}, J., {Mankovich}, C.~R., \& {Fuller}, J. 2022, MNRAS, 516, 1, \dodoi{10.1093/mnras/stac1957}
    
    \bibitem[{{Dewberry} {et~al.}(2021){Dewberry}, {Mankovich}, {Fuller}, {Lai}, \&
      {Xu}}]{2021PSJ.....2..198D}
    {Dewberry}, J.~W., {Mankovich}, C.~R., {Fuller}, J., {Lai}, D., \& {Xu}, W.
      2021, \psj, 2, 198, \dodoi{10.3847/PSJ/ac0e2a}
    
    \bibitem[{{Durante} {et~al.}(2022){Durante}, {Guillot}, {Iess}, {Stevenson},
      {Mankovich}, {Markham}, {Galanti}, {Kaspi}, {Zannoni}, {Gomez Casajus},
      {Lari}, {Parisi}, {Buccino}, {Park}, \& {Bolton}}]{2022NatCo..13.4632D}
    {Durante}, D., {Guillot}, T., {Iess}, L., {et~al.} 2022, Nature Communications,
      13, 4632, \dodoi{10.1038/s41467-022-32299-9}
    
    \bibitem[{{Fletcher} {et~al.}(2009){Fletcher}, {Orton}, {Teanby}, {Irwin}, \&
      {Bjoraker}}]{2009Icar..199..351F}
    {Fletcher}, L.~N., {Orton}, G.~S., {Teanby}, N.~A., {Irwin}, P.~G.~J., \&
      {Bjoraker}, G.~L. 2009, \icarus, 199, 351,
      \dodoi{10.1016/j.icarus.2008.09.019}
    
    \bibitem[{{Foreman-Mackey} {et~al.}(2013){Foreman-Mackey}, {Hogg}, {Lang}, \&
      {Goodman}}]{2013PSP..125..306F}
    {Foreman-Mackey}, D., {Hogg}, D.~W., {Lang}, D., \& {Goodman}, J. 2013, \pasp,
      125, 306, \dodoi{10.1086/670067}
    
    \bibitem[{{French} {et~al.}(2021){French}, {Bridges}, {Hedman}, {Nicholson},
      {Mankovich}, \& {McGhee-French}}]{2021Icar..37014660F}
    {French}, R.~G., {Bridges}, B., {Hedman}, M.~M., {et~al.} 2021, \icarus, 370,
      114660, \dodoi{10.1016/j.icarus.2021.114660}
    
    \bibitem[{{French} {et~al.}(2019){French}, {McGhee-French}, {Nicholson}, \&
      {Hedman}}]{2019Icar..319..599F}
    {French}, R.~G., {McGhee-French}, C.~A., {Nicholson}, P.~D., \& {Hedman}, M.~M.
      2019, \icarus, 319, 599, \dodoi{10.1016/j.icarus.2018.10.013}
    
    \bibitem[{{French} {et~al.}(2016){French}, {Nicholson}, {Hedman}, {Hahn},
      {McGhee-French}, {Colwell}, {Marouf}, \& {Rappaport}}]{2016Icar..279...62F}
    {French}, R.~G., {Nicholson}, P.~D., {Hedman}, M.~M., {et~al.} 2016, \icarus,
      279, 62, \dodoi{10.1016/j.icarus.2015.08.020}
    
    \bibitem[{{Fuller}(2014)}]{2014Icar..242..283F}
    {Fuller}, J. 2014, \icarus, 242, 283, \dodoi{10.1016/j.icarus.2014.08.006}
    
    \bibitem[{{Galanti} \& {Kaspi}(2017)}]{2017ApJ...843L..25G}
    {Galanti}, E., \& {Kaspi}, Y. 2017, \apjl, 843, L25,
      \dodoi{10.3847/2041-8213/aa7aec}
    
    \bibitem[{{Galanti} \& {Kaspi}(2021)}]{2021MNRAS.501.2352G}
    ---. 2021, \mnras, 501, 2352, \dodoi{10.1093/mnras/staa3722}
    
    \bibitem[{{Galanti} {et~al.}(2019){Galanti}, {Kaspi}, {Miguel}, {Guillot},
      {Durante}, {Racioppa}, \& {Iess}}]{2019GeoRL..46..616G}
    {Galanti}, E., {Kaspi}, Y., {Miguel}, Y., {et~al.} 2019, \grl, 46, 616,
      \dodoi{10.1029/2018GL078087}
    
    \bibitem[{{Garc{\'\i}a-Melendo} {et~al.}(2011){Garc{\'\i}a-Melendo},
      {P{\'e}rez-Hoyos}, {S{\'a}nchez-Lavega}, \& {Hueso}}]{2011Icar..215...62G}
    {Garc{\'\i}a-Melendo}, E., {P{\'e}rez-Hoyos}, S., {S{\'a}nchez-Lavega}, A., \&
      {Hueso}, R. 2011, \icarus, 215, 62, \dodoi{10.1016/j.icarus.2011.07.005}
    
    \bibitem[{{Giampieri} {et~al.}(2006){Giampieri}, {Dougherty}, {Smith}, \&
      {Russell}}]{2006Natur.441...62G}
    {Giampieri}, G., {Dougherty}, M.~K., {Smith}, E.~J., \& {Russell}, C.~T. 2006,
      \nat, 441, 62, \dodoi{10.1038/nature04750}
    
    \bibitem[{{Guillot}(1999)}]{1999P&SS...47.1183G}
    {Guillot}, T. 1999, \planss, 47, 1183, \dodoi{10.1016/S0032-0633(99)00043-4}
    
    \bibitem[{{Guillot} {et~al.}(2022){Guillot}, {Fletcher}, {Helled}, {Ikoma},
      {Line}, \& {Parmentier}}]{2022arXiv220504100G}
    {Guillot}, T., {Fletcher}, L.~N., {Helled}, R., {et~al.} 2022, arXiv e-prints,
      arXiv:2205.04100.
    \newblock \doarXiv{2205.04100}
    
    \bibitem[{{Guillot} {et~al.}(2018){Guillot}, {Miguel}, {Militzer}, {Hubbard},
      {Kaspi}, {Galanti}, {Cao}, {Helled}, {Wahl}, {Iess}, {Folkner}, {Stevenson},
      {Lunine}, {Reese}, {Biekman}, {Parisi}, {Durante}, {Connerney}, {Levin}, \&
      {Bolton}}]{2018Natur.555..227G}
    {Guillot}, T., {Miguel}, Y., {Militzer}, B., {et~al.} 2018, \nat, 555, 227,
      \dodoi{10.1038/nature25775}
    
    \bibitem[{{Hedman} \& {Nicholson}(2013)}]{2013AJ....146...12H}
    {Hedman}, M.~M., \& {Nicholson}, P.~D. 2013, \aj, 146, 12,
      \dodoi{10.1088/0004-6256/146/1/12}
    
    \bibitem[{{Hedman} \& {Nicholson}(2014)}]{2014MNRAS.444.1369H}
    ---. 2014, \mnras, 444, 1369, \dodoi{10.1093/mnras/stu1503}
    
    \bibitem[{{Hedman} {et~al.}(2022){Hedman}, {Nicholson}, {El Moutamid}, \&
      {Smotherman}}]{2022PSJ.....3...61H}
    {Hedman}, M.~M., {Nicholson}, P.~D., {El Moutamid}, M., \& {Smotherman}, S.
      2022, \psj, 3, 61, \dodoi{10.3847/PSJ/ac4df8}
    
    \bibitem[{{Hedman} {et~al.}(2019){Hedman}, {Nicholson}, \&
      {French}}]{2019AJ....157...18H}
    {Hedman}, M.~M., {Nicholson}, P.~D., \& {French}, R.~G. 2019, \aj, 157, 18,
      \dodoi{10.3847/1538-3881/aaf0a6}
    
    \bibitem[{{Helled} {et~al.}(2010){Helled}, {Anderson}, \&
      {Schubert}}]{2010Icar..210..446H}
    {Helled}, R., {Anderson}, J.~D., \& {Schubert}, G. 2010, \icarus, 210, 446,
      \dodoi{10.1016/j.icarus.2010.06.037}
    
    \bibitem[{{Helled} {et~al.}(2015){Helled}, {Galanti}, \&
      {Kaspi}}]{2015Natur.520..202H}
    {Helled}, R., {Galanti}, E., \& {Kaspi}, Y. 2015, \nat, 520, 202,
      \dodoi{10.1038/nature14278}
    
    \bibitem[{{Helled} {et~al.}(2020){Helled}, {Mazzola}, \&
      {Redmer}}]{2020NatRP...2..562H}
    {Helled}, R., {Mazzola}, G., \& {Redmer}, R. 2020, Nature Reviews Physics, 2,
      562, \dodoi{10.1038/s42254-020-0223-3}
    
    \bibitem[{{Hubbard}(2013)}]{2013ApJ...768...43H}
    {Hubbard}, W.~B. 2013, \apj, 768, 43, \dodoi{10.1088/0004-637X/768/1/43}
    
    \bibitem[{{Iess} {et~al.}(2019){Iess}, {Militzer}, {Kaspi}, {Nicholson},
      {Durante}, {Racioppa}, {Anabtawi}, {Galanti}, {Hubbard}, {Mariani},
      {Tortora}, {Wahl}, \& {Zannoni}}]{2019Sci...364.2965I}
    {Iess}, L., {Militzer}, B., {Kaspi}, Y., {et~al.} 2019, Science, 364, aat2965,
      \dodoi{10.1126/science.aat2965}
    
    \bibitem[{{Jacobson}(2022)}]{2022AJ....164..199J}
    {Jacobson}, R.~A. 2022, \aj, 164, 199, \dodoi{10.3847/1538-3881/ac90c9}
    
    \bibitem[{{Kaspi} {et~al.}(2018){Kaspi}, {Galanti}, {Hubbard}, {Stevenson},
      {Bolton}, {Iess}, {Guillot}, {Bloxham}, {Connerney}, {Cao}, {Durante},
      {Folkner}, {Helled}, {Ingersoll}, {Levin}, {Lunine}, {Miguel}, {Militzer},
      {Parisi}, \& {Wahl}}]{2018Natur.555..223K}
    {Kaspi}, Y., {Galanti}, E., {Hubbard}, W.~B., {et~al.} 2018, \nat, 555, 223,
      \dodoi{10.1038/nature25793}
    
    \bibitem[{{Koskinen} \& {Guerlet}(2018)}]{2018Icar..307..161K}
    {Koskinen}, T.~T., \& {Guerlet}, S. 2018, \icarus, 307, 161,
      \dodoi{10.1016/j.icarus.2018.02.020}
    
    \bibitem[{{Lindal} {et~al.}(1985){Lindal}, {Sweetnam}, \&
      {Eshleman}}]{1985AJ.....90.1136L}
    {Lindal}, G.~F., {Sweetnam}, D.~N., \& {Eshleman}, V.~R. 1985, \aj, 90, 1136,
      \dodoi{10.1086/113820}
    
    \bibitem[{{Liu} {et~al.}(2008){Liu}, {Goldreich}, \&
      {Stevenson}}]{2008Icar..196..653L}
    {Liu}, J., {Goldreich}, P.~M., \& {Stevenson}, D.~J. 2008, \icarus, 196, 653,
      \dodoi{10.1016/j.icarus.2007.11.036}
    
    \bibitem[{{Lodders}(2003)}]{2003ApJ...591.1220L}
    {Lodders}, K. 2003, \apj, 591, 1220, \dodoi{10.1086/375492}
    
    \bibitem[{{Mankovich} {et~al.}(2019){Mankovich}, {Marley}, {Fortney}, \&
      {Movshovitz}}]{2019ApJ...871....1M}
    {Mankovich}, C., {Marley}, M.~S., {Fortney}, J.~J., \& {Movshovitz}, N. 2019,
      \apj, 871, 1, \dodoi{10.3847/1538-4357/aaf798}
    
    \bibitem[{{Mankovich} \& {Fuller}(2021)}]{2021NatAs...5.1103M}
    {Mankovich}, C.~R., \& {Fuller}, J. 2021, Nature Astronomy, 5, 1103,
      \dodoi{10.1038/s41550-021-01448-3}
    
    \bibitem[{{Markham} {et~al.}(2020){Markham}, {Durante}, {Iess}, \&
      {Stevenson}}]{2020PSJ.....1...27M}
    {Markham}, S., {Durante}, D., {Iess}, L., \& {Stevenson}, D. 2020, \psj, 1, 27,
      \dodoi{10.3847/PSJ/ab9f21}
    
    \bibitem[{{Marley}(1991)}]{1991Icar...94..420M}
    {Marley}, M.~S. 1991, \icarus, 94, 420, \dodoi{10.1016/0019-1035(91)90239-P}
    
    \bibitem[{{Marley} \& {Porco}(1993)}]{1993Icar..106..508M}
    {Marley}, M.~S., \& {Porco}, C.~C. 1993, \icarus, 106, 508,
      \dodoi{10.1006/icar.1993.1189}
    
    \bibitem[{{Miguel} {et~al.}(2016){Miguel}, {Guillot}, \&
      {Fayon}}]{2016A&A...596A.114M}
    {Miguel}, Y., {Guillot}, T., \& {Fayon}, L. 2016, \aap, 596, A114,
      \dodoi{10.1051/0004-6361/201629732}
    
    \bibitem[{{Miguel} {et~al.}(2022){Miguel}, {Bazot}, {Guillot}, {Howard},
      {Galanti}, {Kaspi}, {Hubbard}, {Militzer}, {Helled}, {Atreya}, {Connerney},
      {Durante}, {Kulowski}, {Lunine}, {Stevenson}, \&
      {Bolton}}]{2022A&A...662A..18M}
    {Miguel}, Y., {Bazot}, M., {Guillot}, T., {et~al.} 2022, \aap, 662, A18,
      \dodoi{10.1051/0004-6361/202243207}
    
    \bibitem[{{Militzer} \& {Hubbard}(2013)}]{2013ApJ...774..148M}
    {Militzer}, B., \& {Hubbard}, W.~B. 2013, \apj, 774, 148,
      \dodoi{10.1088/0004-637X/774/2/148}
    
    \bibitem[{{Militzer} {et~al.}(2019){Militzer}, {Wahl}, \&
      {Hubbard}}]{2019ApJ...879...78M}
    {Militzer}, B., {Wahl}, S., \& {Hubbard}, W.~B. 2019, \apj, 879, 78,
      \dodoi{10.3847/1538-4357/ab23f0}
    
    \bibitem[{{Militzer} {et~al.}(2022){Militzer}, {Hubbard}, {Wahl}, {Lunine},
      {Galanti}, {Kaspi}, {Miguel}, {Guillot}, {Moore}, {Parisi}, {Connerney},
      {Helled}, {Cao}, {Mankovich}, {Stevenson}, {Park}, {Wong}, {Atreya},
      {Anderson}, \& {Bolton}}]{BM22}
    {Militzer}, B., {Hubbard}, W.~B., {Wahl}, S., {et~al.} 2022, \psj, 3, 185,
      \dodoi{10.3847/PSJ/ac7ec8}
    
    \bibitem[{{Moore} {et~al.}(2021){Moore}, {Bolton}, {Cao}, {K. Dougherty}, \&
      {Bloxham}}]{2021PSJ.....2..181M}
    {Moore}, K.~M., {Bolton}, B., {Cao}, H., {K. Dougherty}, M., \& {Bloxham}, J.
      2021, \psj, 2, 181, \dodoi{10.3847/PSJ/ac173c}
    
    \bibitem[{{Movshovitz} \& {Fortney}(2022)}]{2022PSJ.....3...88M}
    {Movshovitz}, N., \& {Fortney}, J.~J. 2022, \psj, 3, 88,
      \dodoi{10.3847/PSJ/ac60ff}
    
    \bibitem[{{Movshovitz} {et~al.}(2020){Movshovitz}, {Fortney}, {Mankovich},
      {Thorngren}, \& {Helled}}]{2020ApJ...891..109M}
    {Movshovitz}, N., {Fortney}, J.~J., {Mankovich}, C., {Thorngren}, D., \&
      {Helled}, R. 2020, \apj, 891, 109, \dodoi{10.3847/1538-4357/ab71ff}
    
    \bibitem[{{Nettelmann} {et~al.}(2021){Nettelmann}, {Movshovitz}, {Ni},
      {Fortney}, {Galanti}, {Kaspi}, {Helled}, {Mankovich}, \&
      {Bolton}}]{2021PSJ.....2..241N}
    {Nettelmann}, N., {Movshovitz}, N., {Ni}, D., {et~al.} 2021, \psj, 2, 241,
      \dodoi{10.3847/PSJ/ac390a}
    
    \bibitem[{{Ouazzani} {et~al.}(2012){Ouazzani}, {Dupret}, \&
      {Reese}}]{2012A&A...547A..75O}
    {Ouazzani}, R.~M., {Dupret}, M.~A., \& {Reese}, D.~R. 2012, \aap, 547, A75,
      \dodoi{10.1051/0004-6361/201219548}
    
    \bibitem[{{Read} {et~al.}(2009){Read}, {Dowling}, \&
      {Schubert}}]{2009Natur.460..608R}
    {Read}, P.~L., {Dowling}, T.~E., \& {Schubert}, G. 2009, \nat, 460, 608,
      \dodoi{10.1038/nature08194}
    
    \bibitem[{{Reese} {et~al.}(2006){Reese}, {Ligni{\`e}res}, \&
      {Rieutord}}]{2006A&A...455..621R}
    {Reese}, D., {Ligni{\`e}res}, F., \& {Rieutord}, M. 2006, \aap, 455, 621,
      \dodoi{10.1051/0004-6361:20065269}
    
    \bibitem[{{Rieutord} {et~al.}(2016){Rieutord}, {Espinosa Lara}, \&
      {Putigny}}]{2016JCoPh.318..277R}
    {Rieutord}, M., {Espinosa Lara}, F., \& {Putigny}, B. 2016, Journal of
      Computational Physics, 318, 277, \dodoi{10.1016/j.jcp.2016.05.011}
    
    \bibitem[{{Rosen} {et~al.}(1991){Rosen}, {Tyler}, {Marouf}, \&
      {Lissauer}}]{1991Icar...93...25R}
    {Rosen}, P.~A., {Tyler}, G.~L., {Marouf}, E.~A., \& {Lissauer}, J.~J. 1991,
      \icarus, 93, 25, \dodoi{10.1016/0019-1035(91)90161-L}
    
    \bibitem[{{Saillenfest} {et~al.}(2021){Saillenfest}, {Lari}, \&
      {Bou{\'e}}}]{Saillenfest2021}
    {Saillenfest}, M., {Lari}, G., \& {Bou{\'e}}, G. 2021, Nature Astronomy, 5,
      345, \dodoi{10.1038/s41550-020-01284-x}
    
    \bibitem[{{Stanley}(2010)}]{2010GeoRL..37.5201S}
    {Stanley}, S. 2010, \grl, 37, L05201, \dodoi{10.1029/2009GL041752}
    
    \bibitem[{{Stevenson}(1982)}]{1982GApFD..21..113S}
    {Stevenson}, D.~J. 1982, Geophysical and Astrophysical Fluid Dynamics, 21, 113,
      \dodoi{10.1080/03091928208209008}
    
    \bibitem[{{Stevenson}(2020)}]{2020AREPS..48..465S}
    ---. 2020, Annual Review of Earth and Planetary Sciences, 48, 465,
      \dodoi{10.1146/annurev-earth-081619-052855}
    
    \bibitem[{{Takata} \& {Saio}(2013)}]{2013PASJ...65...68T}
    {Takata}, M., \& {Saio}, H. 2013, \pasj, 65, 68, \dodoi{10.1093/pasj/65.3.68}
    
    \bibitem[{{Thompson}(1990)}]{aneos}
    {Thompson}, S.~L. 1990, ANEOS--—Analytic Equations of State for Shock Physics
      Codes, Sandia Natl. Lab. Doc. SAND89-2951,
      \url{http://prod.sandia.gov/techlib/access-control.cgi/1989/892951.pdf}
    
    \bibitem[{{Wahl} {et~al.}(2017){Wahl}, {Hubbard}, {Militzer}, {Guillot},
      {Miguel}, {Movshovitz}, {Kaspi}, {Helled}, {Reese}, {Galanti}, {Levin},
      {Connerney}, \& {Bolton}}]{2017GeoRL..44.4649W}
    {Wahl}, S.~M., {Hubbard}, W.~B., {Militzer}, B., {et~al.} 2017, \grl, 44, 4649,
      \dodoi{10.1002/2017GL073160}
    
    \bibitem[{{Ward} \& {Hamilton}(2004)}]{2004AJ....128.2501W}
    {Ward}, W.~R., \& {Hamilton}, D.~P. 2004, \aj, 128, 2501,
      \dodoi{10.1086/424533}
    
    \bibitem[{Wisdom {et~al.}(2022)Wisdom, Dbouk, Militzer, Hubbard, Nimmo, Downey,
      \& French}]{doi:10.1126/science.abn1234}
    Wisdom, J., Dbouk, R., Militzer, B., {et~al.} 2022, Science, 377, 1285,
      \dodoi{10.1126/science.abn1234}
    
    \bibitem[{{Wisdom} \& {Hubbard}(2016)}]{2016Icar..267..315W}
    {Wisdom}, J., \& {Hubbard}, W.~B. 2016, \icarus, 267, 315,
      \dodoi{10.1016/j.icarus.2015.12.030}
    
    \bibitem[{{Xu} \& {Lai}(2017)}]{2017PhRvD..96h3005X}
    {Xu}, W., \& {Lai}, D. 2017, \prd, 96, 083005,
      \dodoi{10.1103/PhysRevD.96.083005}
    
    \bibitem[{{Yan} \& {Stanley}(2021)}]{2021AGUA....200318Y}
    {Yan}, C., \& {Stanley}, S. 2021, AGU Advances, 2, e00318,
      \dodoi{10.1029/2020AV000318}
    
    \bibitem[{{Ye} {et~al.}(2018){Ye}, {Fischer}, {Kurth}, {Menietti}, \&
      {Gurnett}}]{2018GeoRL..45.7297Y}
    {Ye}, S.~Y., {Fischer}, G., {Kurth}, W.~S., {Menietti}, J.~D., \& {Gurnett},
      D.~A. 2018, \grl, 45, 7297, \dodoi{10.1029/2018GL077976}
    
    \bibitem[{{Zhang} {et~al.}(2015){Zhang}, {Kong}, \&
      {Schubert}}]{2015ApJ...806..270Z}
    {Zhang}, K., {Kong}, D., \& {Schubert}, G. 2015, \apj, 806, 270,
      \dodoi{10.1088/0004-637X/806/2/270}
    
    \bibitem[{{Zharkov} \& {Trubitsyn}(1978)}]{1978ppi..book.....Z}
    {Zharkov}, V.~N., \& {Trubitsyn}, V.~P. 1978, {Physics of Planetary Interiors}
      (Tucson, AZ: Pachart)
    
    \end{thebibliography}

\appendix
\allowdisplaybreaks
\section{Numerical method for calculating $\delta J_{\lowercase{n}}$}\label{app:tgwe}
Consider a rigidly rotating equilibrium state characterized by a density $\rho_0$, pressure $P_0,$ velocity field ${\bf u}_0={\bf\Omega}_0\times{\bf r}$ (${\bf \Omega}_0=\Omega_0\hat{\bf z}$), and gravity field $\Phi_0$ that satisfy the steady-state Euler and Poisson equations. The linearized equation of motion for time-independent, axisymmetric perturbations $\rho'(r,\theta)$, $P'(r,\theta)$, ${\bf u}=r\sin\theta\delta\Omega(r,\theta)\hat{\boldsymbol{\phi}}$, $\Phi'(r,\theta)$ to this rigidly rotating equilibrium is
\begin{equation}\label{eq:linEoM}
    {\bf G}\rho' - \nabla P' - \rho_0\nabla\Phi'
    =2\boldsymbol{\Omega}_0\times(\rho_0{\bf u}),
\end{equation}
where ${\bf G}=\rho_0^{-1}\nabla P_0$ is the ``effective gravity'' associated with the rigidly rotating state. The barotropic nature of the rigidly rotating equilibrium implies that $\nabla\times {\bf G}=0,$ so taking the curl of \autoref{eq:linEoM} produces
\begin{equation}\label{eq:tgweVec}
    {\bf G}\times\nabla\rho'
    +\nabla\rho_0\times\nabla\Phi'
    =2{\bf \Omega}_0\cdot\nabla(\rho_0{\bf u}).
\end{equation}
$\rho'$ and $\Phi'$ are additionally related by $\nabla^2\Phi'=4\pi G\rho'$. \autoref{eq:tgweVec} can be expressed on the covariant basis of an arbitrary curvilinear coordinate system as
\begin{equation}
    \epsilon^{ijk}[
        G_j\partial_k\rho'
        +\partial_j\rho_0\partial_k\Phi'
    ]
    =2\Omega_0^j[
        \partial_j(\rho_0u^i)
        +\rho_0\Gamma^i_{kj}u^k
    ],
\end{equation}
where paired upper and lower indices denote summation, $\partial_i$ denotes partial differentiation with respect to coordinate $x^i$, $\epsilon^{ijk}=J^{-1}[i,j,k]$ for $J$ the Jacobian of the coordinate system, and $\Gamma^i_{kj}$ are Christoffel symbols. 

As in our mode calculations (see \autoref{sec.methods.modes}), we adopt the non-orthogonal coordinate system $(\zeta,\theta,\phi)$ described in D21 \citep[and originally proposed by][]{1998PhRvD..58j4020B}, which involves a quasi-radial coordinate $\zeta$ constructed so that $\zeta=1$ on the pre-computed surface of the rigidly rotating planet. Retaining nonzero geometric factors \citep[summarized in ][]{2016JCoPh.318..277R}, we find 
\begin{equation}
    (
        G_\zeta\partial_\theta 
        -G_\theta\partial_\zeta
    )\rho'
    +(
        \partial_\zeta\rho_0\partial_\theta 
        -\partial_\theta\rho_0\partial_\zeta
    )\Phi'
    =2\Omega_0r\sin\theta\left[
        (r\cos\theta+r_\theta \sin\theta)\partial_\zeta 
        -r_\zeta \sin\theta\partial_\theta
    \right](\rho_0\delta\Omega)
    :=f(r,\theta),
\end{equation}
where $G_i=\rho_0^{-1}\partial_i P_0,$ and $r_i=\partial_i r$ (note that $r=r(\zeta,\theta)$).

Inserting expansions $\rho'=\sum_n\rho^n(\zeta)Y_n^0(\theta,\phi)$, $\Phi'=\sum_n\Phi^n(\zeta)Y_n^0(\theta,\phi)$ in zonal spherical harmonics, we project onto an arbitrary degree $n=\ell$ by taking the inner product with $\partial_\theta Y_\ell^0$. This leads to an infinite set of coupled ordinary differential equations in $\zeta$:
\begin{equation}\label{eq:tgweProj}
    \sum_n\left[
        \langle \partial_\theta Y_\ell^0,\partial_\theta Y_n^0G_\zeta
        \rangle 
        -\langle \partial_\theta Y_\ell^0,Y_n^0G_\theta 
        \rangle 
        \partial_\zeta
    \right]\rho^n
    +\sum_n\left[
        \langle 
            \partial_\theta Y_\ell^0,\partial_\theta Y_n^0\partial_\zeta\rho_0
        \rangle 
        -\langle 
            \partial_\theta Y_\ell^0, Y_n^0\partial_\theta\rho_0
        \rangle
        \partial_\zeta
    \right]\Phi^n
    =\langle \partial_\theta Y_\ell^0,f\rangle,
\end{equation}
where $\langle \partial_\theta Y_\ell^0,h\rangle=\int_0^\pi\int_0^{2\pi} h\partial_\theta Y_\ell^0\sin\theta\text{d}\phi\text{d}\theta$ for a given function $h(r,\theta)$. 

We solve \autoref{eq:tgweVec} simultaneously with Poisson's equation (the projected form of which is given by eq. A32 in D21), using the same Chebyshev collocation method applied to mode calculations. For boundary conditions, we use boundary bordering to enforce a vanishing gravitational perturbation at infinity. We additionally impose regularity boundary conditions on both $\rho'$ and $\Phi'$ at the origin. 

Lastly, we note that \autoref{eq:tgweProj} does not constrain $\rho^0,$ since the integrals on the lefthand side vanish for $\ell=0$. In a spherical, nonrotating planet, this would have no effect on the gravitational moments \citep[e.g.,][]{2015ApJ...806..270Z}; in a centrifugally flattened planet this is no longer the case, but we find that the higher-degree $\delta J_n$ are still largely insensitive to the prescription for $\rho^0,$ provided that we ensure $\delta J_0=0.$ Since the gravitational moments are directly related to the gravitational perturbation in the external vacuum  \citep[e.g.,][]{D22}, including self-gravity allows for convenient application of this additional constraint. 

\begin{figure}
    \centering
    \includegraphics[width=0.495\columnwidth]{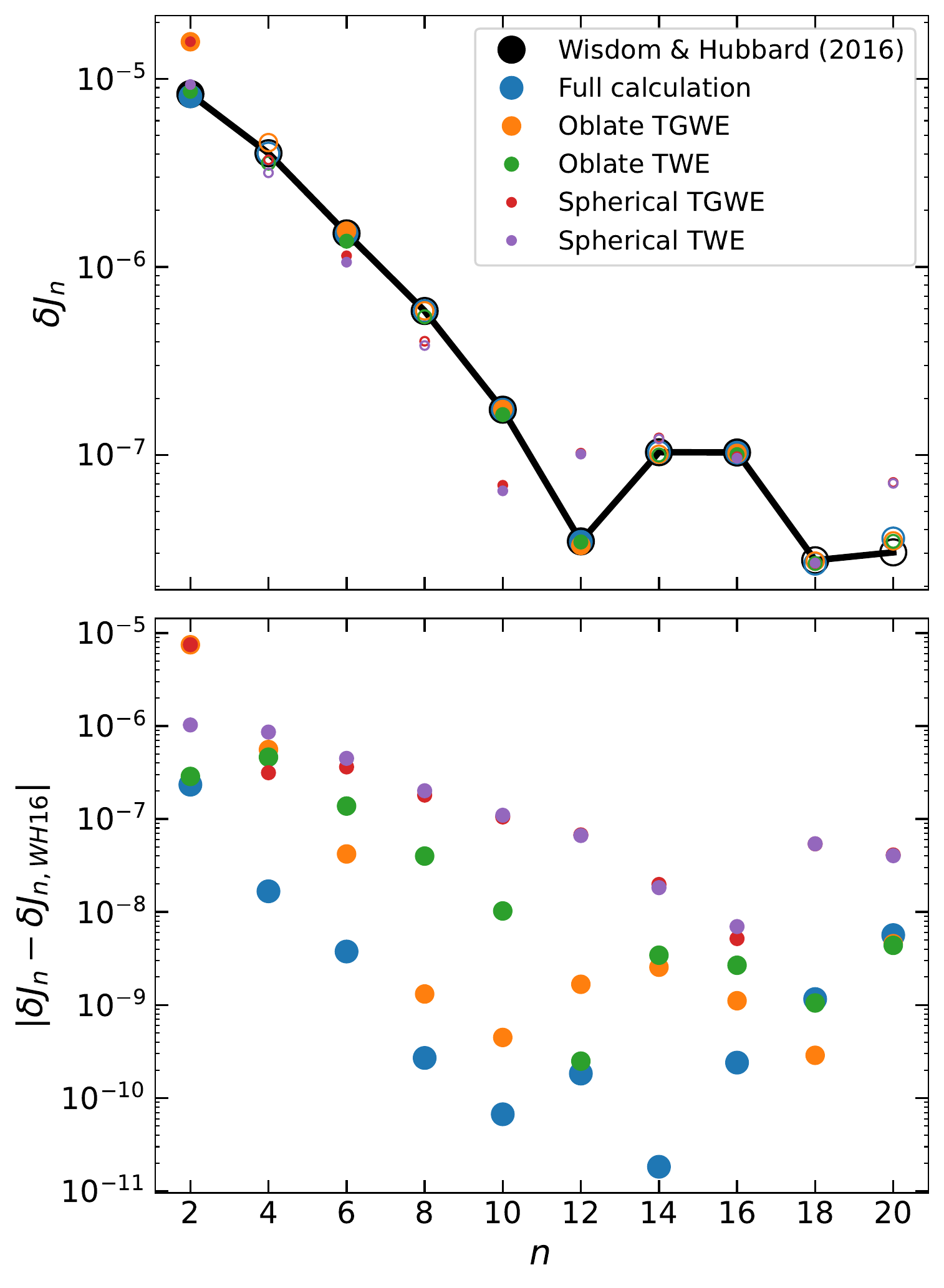}
    \includegraphics[width=0.495\columnwidth]{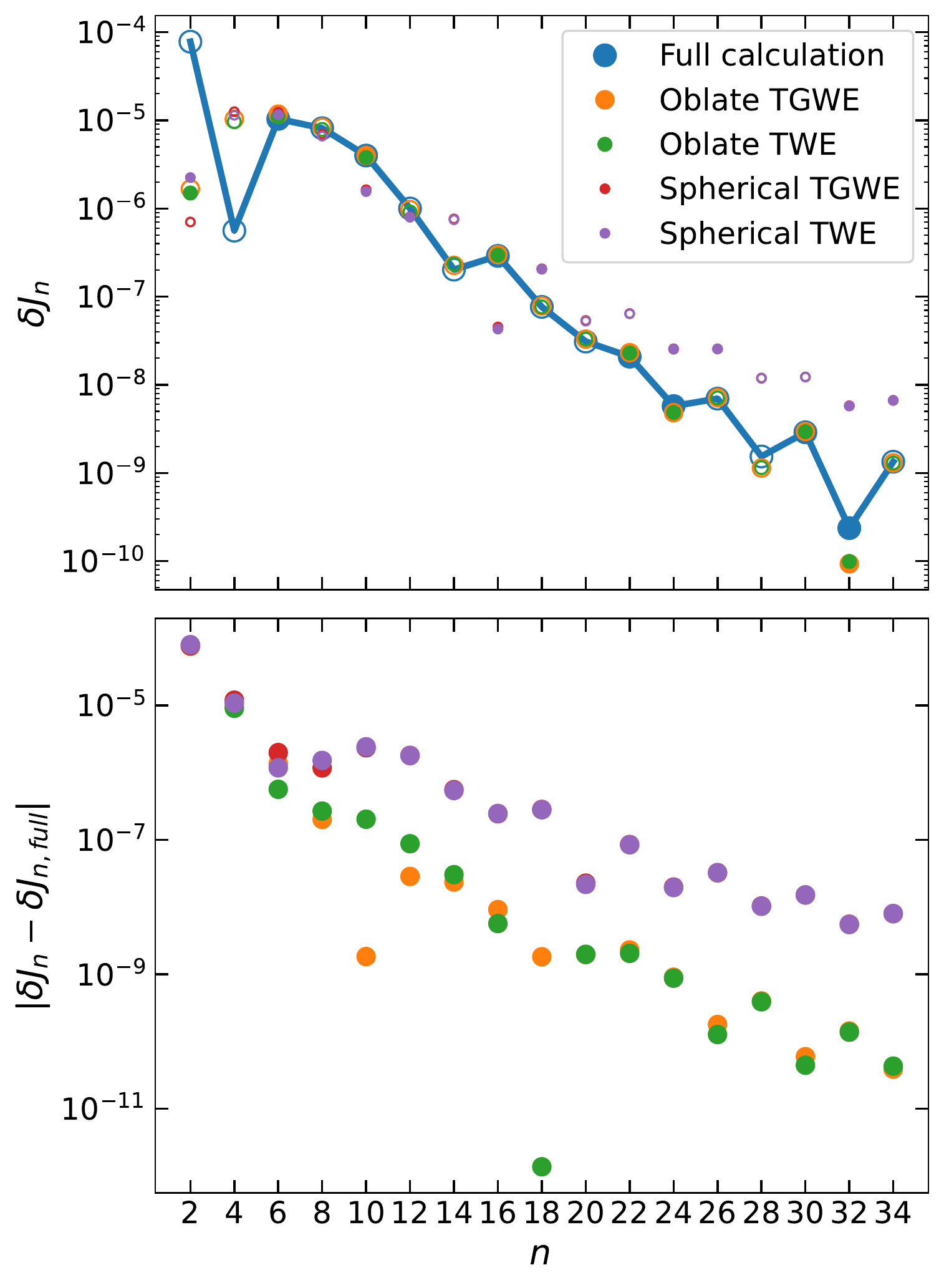}
    \caption{Gravitational coefficient shifts $\delta J_n$ (top) and associated residuals (bottom), computed for the ``DR3'' differential rotation profile considered by \cite{2016Icar..267..315W} (left) and one of the differential rotation profiles from \cite{D22} (right). Differently colored points show $\delta J_n$ calculated directly from the Euler and Poisson equations (black, blue), and from the TGWE/TWE under various approximations (orange, green, red, purple).}
    \label{fig:tgwe}
\end{figure}

The lefthand panels in \autoref{fig:tgwe} compare the results of calculations under various approximations against $\delta J_n$ reported by \cite{2016Icar..267..315W} for their ``DR3'' differential rotation profile. The top panel shows raw $\delta J_n$ calculations, with negative values indicated by empty points, while the bottom panel shows residuals from the calculations of \cite{2016Icar..267..315W}. Blue points indicate results from self-consistent polytropic model calculations of \cite{D22}, and orange points show the result of calculations solving the oblate TGWE (\autoref{eq:tgweProj}) in full (the approach taken in this work). Meanwhile green points show calculations in which we ignore the gravitational perturbation in \autoref{eq:tgweProj}, solving the TWE while still including the geometric effects associated with planetary oblateness. Lastly, the red and purple points respectively indicate solutions of the TGWE and TWE in a spherical geometry. The righthand panels in \autoref{fig:tgwe} show analogous results for one of the differential rotation profiles considered by \cite{D22} (adopting a value $d=0.3$ for their depth parameter). The calculations in the lefthand (righthand) panels were performed with $N_\zeta=80$ ($90$) collocation points in the quasiradial direction, and a spherical harmonic truncation of $N_\ell=30$ ($60$).

In agreement with \cite{2017JGRE..122..686C}, \autoref{fig:tgwe} suggests that the planetary oblateness is more important to include than self-gravity for the higher degree $\delta J_n$, and that $\delta J_2$ and $\delta J_4$ computed from the thermal wind equation under any approximation should be treated with caution. As noted by \cite{2017JGRE..122..686C}, however, uncertainties in $\delta J_2$ and $\delta J_4$ due to winds are insignificant compared with uncertainties resulting from uncertain composition parameters or equations of state.

\section{Effect of omitting the $\lowercase{m}=3,4,5$ constraints}\label{app.omitm345}

{The largest tension between our favored model and the data is in the pattern speeds of the $m=3,4,5$ Lindblad resonances (Figure~\ref{fig.audit_best}) that we attribute to sectoral f-modes. These deep-seated modes are highly sensitive to the Brunt-V\"ais\"al\"a frequency, and hence to the composition profile, within the deep interior. Consequently their large pattern speed residuals ($\approx1-6\ \degd$) likely stem from the model's simplistic composition profile. 
}

\begin{figure}
    \centering
    \includegraphics[width=\columnwidth]{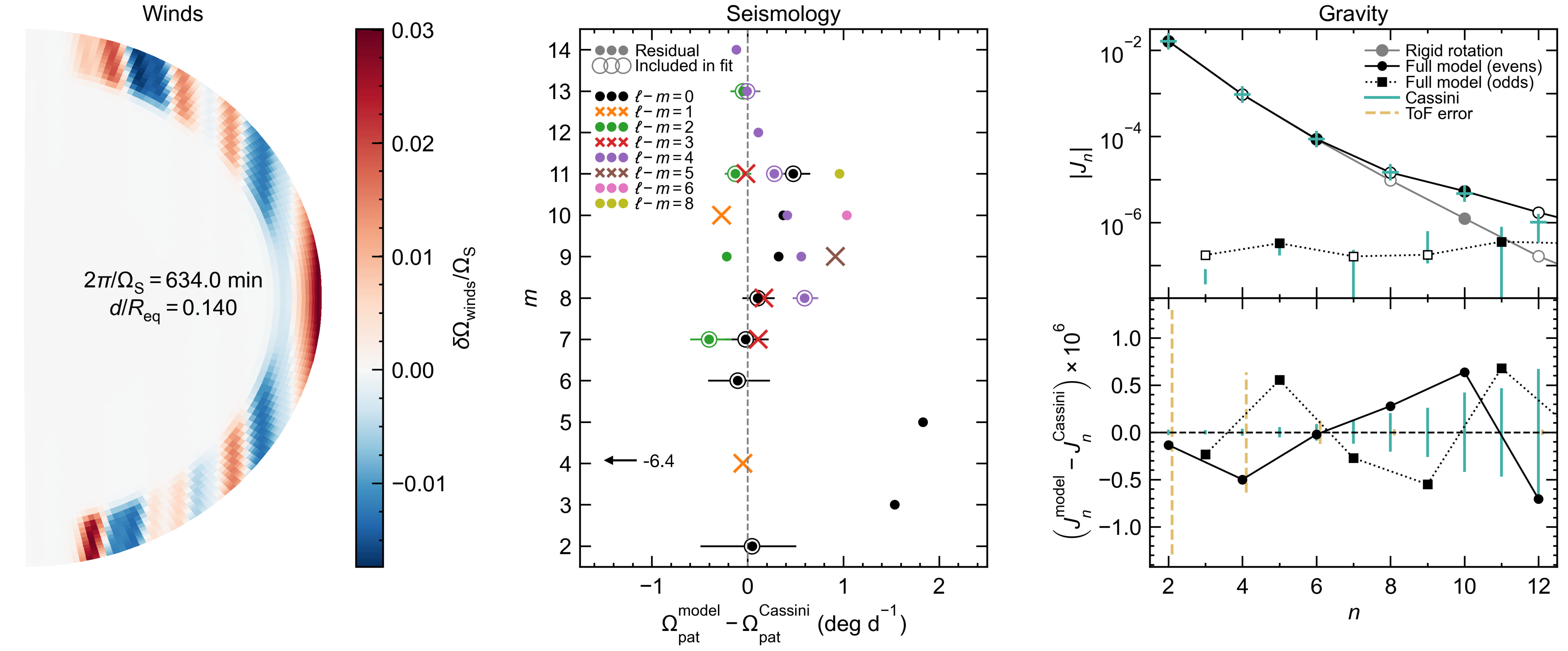}
    \caption{
    \label{fig.omitm345}
    {As in Figure~\ref{fig.audit_best}, but for a sample neglecting the $m=3,4,5$ constraints. Here the central panel includes vertical resonances (crosses) in addition to Lindblad resonances (circles).
    }}
\end{figure}

{To ensure that these less well understood modes, particularly the outlying $m=4$ mode, do not bias our results, we create a sample that neglects the $m=3,4,5$ modes but is otherwise identical to the main ``precise seismology'' calculation in Figures~\ref{fig.audit_best}-\ref{fig.1dpost}.
The best model in the alternate sample is summarized in Figure~\ref{fig.omitm345}. }

{
The exclusion of the $\ell=m=4$ mode in particular, which all of our models strongly underestimate, allows this model to better fit the remaining modes by adopting a modestly faster deep spin (634.0 min versus Figure~\ref{fig.audit_best}'s 634.8 min) and deeper wind decay depth ($0.140\,\req$ versus $0.132\,\req)$. These differences are within the posterior distributions for each parameter from our main calculation (Figure~\ref{fig.1dpost}).
The systematic offsets as a function of $m$ or $\ell-m$ remain virtually unchanged. We conclude that the poorly fit $m=3,4,5$ modes do not fundamentally bias our main results, but we urge that future work prioritize these low-$m$ modes as a window into Saturn's still uncertain composition profile.
}

\end{document}